# Beyond Performance Scores: Directed Functional Connectivity as a Brain-Based Biomarker for Motor Skill Learning and Retention

Anil Kamat[1], Rahul Rahul[1], Lora Cavuoto[2], Harry Burke[4], Matthew Hackett[5], Jack Norfleet[5], Steven Schwaitzberg[3], Suvranu De[6]

[1]Center for Modeling, Simulation, and Imaging for Medicine, Rensselaer Polytechnic Institute, Troy, New York 12180, USA
[2]Department of Industrial and Systems Engineering, University at Buffalo, Buffalo, NY
[3]Department of Surgery, University at Buffalo Jacobs School of Medicine and Biomedical Sciences, Buffalo, NY
[4]Department of Medicine, F. Edward Hébert School of Medicine, Uniformed Services University of the Health Sciences, Bethesda, MD
[5]U.S. Army Futures Command, Combat Capabilities Development Command Soldier Center STTC, Orlando FL
[6]College of Engineering, Florida A&M University-Florida State University, Tallahassee, FL 32310, USA

**Abstract**

Motor skill acquisition in fields like surgery, robotics, and sports involves learning complex task sequences through extensive training. Traditional performance metrics, like execution time and error rates, offer limited insight as they fail to capture the neural mechanisms underlying skill learning and retention. This study introduces directed functional connectivity (dFC), derived from electroencephalography (EEG), as a novel brain-based biomarker for assessing motor skill learning and retention. For the first time, dFC is applied as a biomarker to map the stages of the Fitts and Posner motor learning model, offering new insights into the neural mechanisms underlying skill acquisition and retention. Unlike traditional measures, it captures both the strength and direction of neural information flow, providing a comprehensive understanding of neural adaptations across different learning stages. The analysis demonstrates that dFC can effectively identify and track the progression through various stages of the Fitts and Posner model. Furthermore, its stability over a six-week washout period highlights its utility in monitoring long-term retention. No significant changes in dFC were observed in a control group, confirming that the observed neural adaptations were specific to training and not due to external factors. By offering a granular view of the learning process at the group and individual levels, dFC facilitates the development of personalized, targeted training protocols aimed at enhancing outcomes in fields where precision and long-term retention are critical, such as surgical education. These findings underscore the value of dFC as a robust biomarker that complements traditional performance metrics, providing a deeper understanding of motor skill learning and retention.

## 1. Introduction

Various fields like surgery, robotics, and sports require accurately performing a complex sequence of tasks involving fine motor skills. However, these tasks are challenging and demand extensive training over a long period to achieve proficiency[1][2]. Monitoring the progress of the trainees with reliable and objective metrics can help in gaining insight into their performance, identifying areas of improvement, and understanding the learning rate. Further, metrics are useful in designing effective training protocols with targeted and efficient learning processes. Traditionally, such metrics include execution time[3], reaction time[4], movement variability[5], and speed-accuracy trade-off curves[6]. However, metrics are output measures that help explain **what** happened during a performance—such as how quickly and how accurately the task was completed. While metrics are valuable for providing a snapshot of performance, they fall short of explaining **how** and **why** skill acquisition occurs at the neural level. Contemporary methods, such as the Cumulative Sum (CUSUM) score are more robust in tracking the learning process, but introduce bias through self-referring statistics and subjective parameters which limit their reliability as indicators of skill learning[7][8]. Additionally, verbal reports[9][10] about the learning process are often imprecise and subjective since motor learning is largely an implicit process

[11]. These limitations highlight the need for more comprehensive approaches that can reveal the underlying neural mechanisms driving skill acquisition.

In contrast to traditional metrics, a biomarker provides an indicator of physiological processes that can reveal the underlying mechanisms of motor skill learning. Brain-based biomarkers, such as those derived from neuroimaging, can provide direct insights into neural adaptations, offering a more comprehensive and reliable measure of skill acquisition that could enhance training effectiveness. Motor skill acquisition is primarily driven by the brain's ability to learn and refine movements or sequences of movements through repeated practice, attributed to adaptations[12] in the underlying neural processes, such as neural interactions[13][14][15][16]. These adaptations can be captured by functional connectivity (FC)[17][18][19] of the brain. While undirected FC has been explored as a measure of motor skill learning, it offers only a limited view of the underlying neural interaction mechanisms. To overcome these limitations, we proposed a biomarker based on directed functional connectivity (dFC), measured through electroencephalography (EEG), as developed in our previous study[20]. Unlike undirected FC, dFC quantifies both the strength and direction of information flow between brain regions, offering a more comprehensive assessment of the neural processes underpinning skill acquisition. By capturing how information moves throughout the brain, dFC serves as a robust biomarker that helps explain the "how" and "why" of skill learning, rather than merely documenting what happened. This deeper understanding of the underlying neural mechanisms can inform the development of targeted training protocols that align with specific neural adaptations, ultimately enhancing learning outcomes.

By using brain-based biomarkers like dFC, we can gain a deeper understanding of the neural processes involved across different stages of motor skill acquisition[21][22]. Fitts and Posner[23][24] proposed a widely recognized model of skill acquisition based on performance scores that centers on three distinct stages: cognitive, associative, and autonomous. While these stages are typically defined by observable performance outcomes, biomarkers like dFC allow us to refine and deepen our understanding of these transitions by directly capturing the neural adaptations occurring at each stage. This provides a more precise and nuanced insight into how skills are learned and retained, going beyond what traditional metrics can offer.

Based on the Fitts and Posner model, the cognitive stage is characterized by the learner's initial attempts to understand the task goals and determine the appropriate sequence of actions At this stage, neural adaptations reflect the reliance on explicit knowledge as learners consciously analyze steps and strategies. Performance at this stage is often slow and prone to errors as the learner becomes familiar with the basic requirements of the task. As the fundamental action sequence is internalized, learners enter the associative stage, where neural adaptations shift toward fine-tuning subparts of the task and optimizing transitions between them. This stage involves refining movements and reducing mistakes, although the conscious effort is still evident. Finally, the autonomous stage marks the routinization of the action, where performance becomes more automated and efficient. Brain-based metrics like dFC provide a unique lens through which we can track these transitions, offering the potential for tailoring training protocols to the learner's neural state at each stage of skill acquisition. While the Fitts and Posner model discusses the neural process associated with each learning stage, it has been mostly explored with performance-based metrics. To our knowledge, this is the first study using a brain-based biomarker (dFC) in the identification of motor learning stages in the Fitts and Posner model.

The primary goal of motor skill learning extends beyond achieving task proficiency to ensuring long-term retention[25], which can range from several weeks to several years[26][27]. dFC can serve as a promising biomarker for assessing both the retention of motor skills and the neural interaction mechanisms that adapt during this process. Motor skill learning occurs across two distinct phases during each training session: active practice and consolidation[28][29] during rest or sleep[30][2]. During active practice, neural interactions are dynamically modified as the brain encodes the new motor task. However, these neural interactions are often fragile and vulnerable to interference. The consolidation phase plays a crucial role in transforming these fragile interactions into robust and stable configurations[31][32], enabling long-term preservation of learned skills.

This stabilization of experience-driven neural adaptations facilitates the retention of task accuracy, even after extended periods without practice[33][34]. During consolidation, procedural memories are transferred and stored predominantly in the motor cortices, such as the left/right primary motor cortices[35] and the supplementary motor area (SMA), reducing reliance on higher cognitive regions. The effectiveness of consolidation is also influenced by the structure of training; distributed practice, involving training spread over several days, has been shown to be more effective for inducing long-term retention than massed training within condensed time frames [27][36]. Conversely, insufficient rest[37] or weak consolidation can result in skill decay or forgetting. While the retention of motor skills has been studied using undirected functional connectivity[38][39][40][41] or brain activation, it remains unexplored with dFC. In this study, we applied dFC to analyze the retention of surgical motor skills[42].

In conclusion, traditional metrics such as speed, accuracy, and efficiency provide valuable information regarding motor skill acquisition but are inherently limited to observable performance outcomes. These metrics fail to reveal the underlying neural adaptations that support the learning and retention of skills. Biomarkers, such as dFC, serve as complementary tools that extend beyond what traditional metrics can provide. They offer critical insights into the neural mechanisms underlying learning, capturing the dynamic neural adaptations that underpin skill acquisition. This dual approach—using metrics to quantify *what* occurs during performance and biomarkers to understand the *how* and *why*—enables a more comprehensive understanding of the learning process. By applying dFC as a biomarker, this study enhances our understanding of the neurophysiological processes associated with skill acquisition and retention, bridging an important gap left by traditional measures. Such an integrated framework allows for the development of individualized, brain-based training protocols aimed at optimizing motor learning and retention. The implications of this approach are particularly profound for fields such as surgical education, where precision, adaptability, and long-term retention are essential to ensuring high levels of expertise and performance.

## 2. Methods

### 2.1. Experimental design

To evaluate our proposed biomarker, we selected a complex bimanual motor task from the surgical domain: the suturing with intracorporeal knot tying task[43]. This task is a critical component of the Fundamentals of Laparoscopic Surgery (FLS) program, a prerequisite for board certification in general surgery[44][45]. The FLS suturing task's complexity and standardization make it an ideal candidate for assessing the efficacy of our novel biomarker in tracking surgical skill acquisition and retention. The task requires placing a 15cm suture through two marks on a Penrose drain model and tying three throws of a knot intracorporeally to close the slit on the Penrose drain. The first knot is the surgeon's knot (i.e. double throw knot) followed by two single throws. The participants need to alternate hands between each throw. After securing the knot, the participants must cut both ends of the suture. The participants are provided with two-needle drivers and a pair of endoscopic scissors to complete the task. The task is performed within an FLS trainer box, with participants viewing their actions on a 2D screen (see Figure 3 for experimental setup).

In this study, 28 students from health-related disciplines, all without prior experience in the FLS laparoscopic suturing with intracorporeal knot-tying task, were recruited for a 15-day training program which was approved by the Institutional Review Board at the university at Buffalo, NY. Participants received detailed instructions on each subtask and were informed of the criteria for successful task completion. The number of trials for each participant increased with each week of training (3 per session for days 1-5, 5 per session for days 6-10, and 7 per session for days 11-15), as participants became more adept in task performance. The block diagram of the study design is shown in Figure 1. The maximum allowed time for each trial was 600 seconds and the trials were separated by a 120 second long rest period. Note that participants completed the task at varying times due to differences in speed. Here, we consider day 1 as pre-training and day 15 as post-training (Figure 1). After six weeks without training, participants returned to perform three trials of the retention task. A second group of 27 participants was recruited as a control group who did not undergo the training program. Instead, they completed two sessions (pre- and post-training) that each consisted of three trials of the FLS suturing task. The demographics of the participants are provided in S. Table 1 and 2. The performance of each trial was evaluated by a proctor using a propriety formula obtained from the FLS committee under a Memorandum of Understanding (MOU) to generate an FLS score.

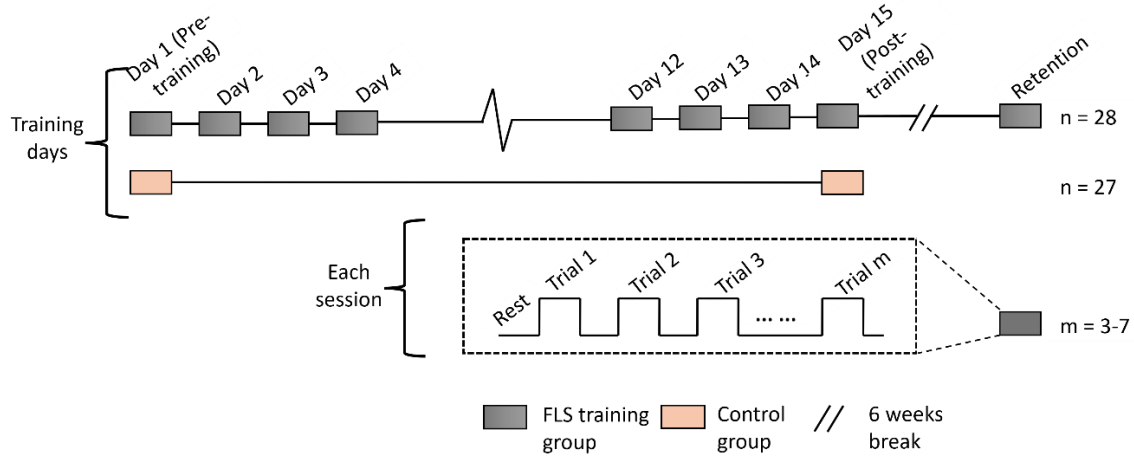

**Figure 1:** The block diagram of the study design is depicted in a gray box chain format, where each box represents a day of the study. The untrained control group (orange) performed the task on the pre-train and again on the post-train day. Note that, "n" is the number of participants and "m" is the number of trails performed by each participant on a single training session.

Further, in this study, we have utilized data from expert participants that were collected in our previous study[20]. The expert group consisted of fifteen participants with an average of 7(±6) years of experience in laparoscopic surgery. They completed three trials of the above-described suturing task on a pre-training day. Additional details about the expert participants can be found in our previous study[20].

2.2. Hardware and equipment

We used a 32-channel wireless LiveAmp system (Brain Vision, USA) EEG montage to record the brain activity of the participants while they performed the suturing and intracorporal knot-tying task (refer to Figure 3 for the experimental setup). EEG is a neuroimaging tool with high temporal resolution, non-invasiveness, and portability. The brain signals were obtained at a sampling frequency of 500 Hz using active gel electrodes. The electrode placement can be seen in the montage in Figure 2. The electrode nomenclature follows the standard 10-5 system[46]. Odd numbers in the figure represent electrodes on the left hemisphere, and even numbers represent electrodes on the right hemisphere. EEG operates by continuously monitoring the postsynaptic potential generated by millions of neurons, offering millisecond-level resolution and direct insight into neural circuit operations. Synchronous neuronal activity generates electrical waves detectable by scalp electrodes, forming rhythmic EEG patterns or oscillations. The oscillations can be categorized into delta ($\delta$; 0.5–3 Hz), theta ($\theta$; 4–7 Hz), alpha ($\alpha$; 8–13 Hz), beta ($\beta$; 14–30 Hz), and gamma ($\gamma$; 30–50 Hz) frequencies. Notably, alpha oscillations, the dominant frequency observed in adult scalp EEG, were utilized in this study due to their well-established association with both cognitive[47] and motor processes[48][49] [50][51].

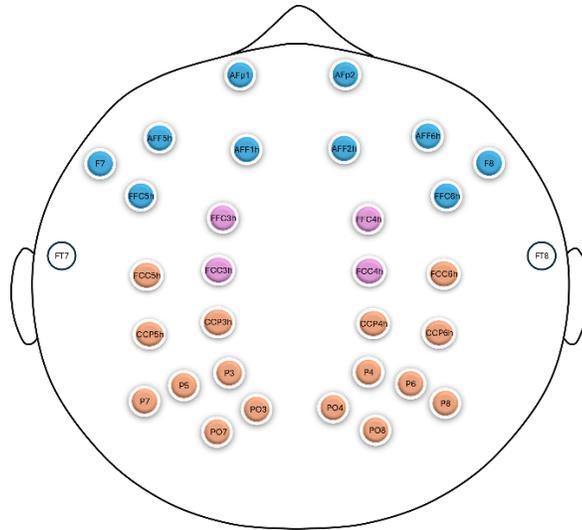

**Figure 2:** EEG montage used to measure the brain activation

## 2.3. Preprocessing

The recorded EEG signals were preprocessed and analyzed offline. The open-source EEGLab toolbox[52] (https://sccn.ucsd.edu/eeglab/index.php), implemented in MATLAB, was used to preprocess the raw brain activation signals to remove the artifacts. The signals were first down sampled to 250Hz, and a high pass filter at 1Hz was applied to remove linear trends or signal drift, and electric interference was removed at 60Hz. Three approaches were used to eliminate bad channels from the data. First, flat channels were removed. Secondly, channels with significant noise were identified and removed based on their standard deviation. Lastly, channels that exhibit poor correlation with other channels were removed using a rejection threshold set at 0.8 for channel correlation. The channels, if removed, were interpolated using neighboring channels' information and spherical spline interpolation. The average reference was computed by subtracting the average of all electrodes from each channel. The current source density (CSD) of the cortical oscillators was computed using eLORETA software to enhance the spatial resolution and reduce the influence of distant sources[53][54].

## 2.4. Brain regions

Several studies have shown the involvement of prefrontal, primary motor, and supplementary motor brain regions and the interactions between them are critical in learning motor skills[1][55][56][57][58][59][39]. Thus, we selected the left/right prefrontal motor cortex (L/R-PFC), the left/right primary motor cortex (L/R-PMC), and the supplementary motor area (SMA) to analyze the neural interaction between them in this study. The 30 channels/electrodes, as shown in the montage (Figure 2), were grouped into 5 distinct brain regions of interest (ROI) as follows: the left prefrontal cortex (LPFC channels: AFp1, AFF5h, F7, AFF1h, and FFC5h), right prefrontal cortex (RPFC channels: AFp2, AFF6H, F8, AFF2h, and FFC6h), supplementary motor area (SMA channels: FFC3h, FCC3h, FFC4h, and FCC4h), left primary motor cortex (LPMC channels: FCC5h, CCP5h, CCP3h, P3, P5, PO7, P7, and PO3), and right primary motor cortex (RPMC channels: FCC6h, CCP6h, CCP4h, P4, P6, PO8, P8, and PO4). All possible connectivity pairs between these brain regions were selected for analysis, resulting in a total of twenty pairs (S. Table 3). These connectivity pairs were used as features for the deep learning model, which is detailed in the "1D CNN with Recursive Feature Elimination (RFE)" section below.

## 2.5. Hierarchical task analysis

FLS suturing with intracorporeal knot tying is a complex task, consisting of several manual subtasks, each requiring distinct fine motor skills and neural interaction[60][61]. We leveraged hierarchical task analysis (HTA) to divide the task into thirteen subtasks. The HTA was performed by a subject matter expert referring to the FLS

guidelines[62][63][64][65]. Figure 3 depicts the various subtasks in procedural order (from top to bottom). This fine-grained analysis is valuable for addressing the detailed aspects of the task and providing neurophysiological interpretation of the findings. When a participant skips a subtask, e.g., they performed the task incorrectly, that subtask is excluded from the corresponding analyses.

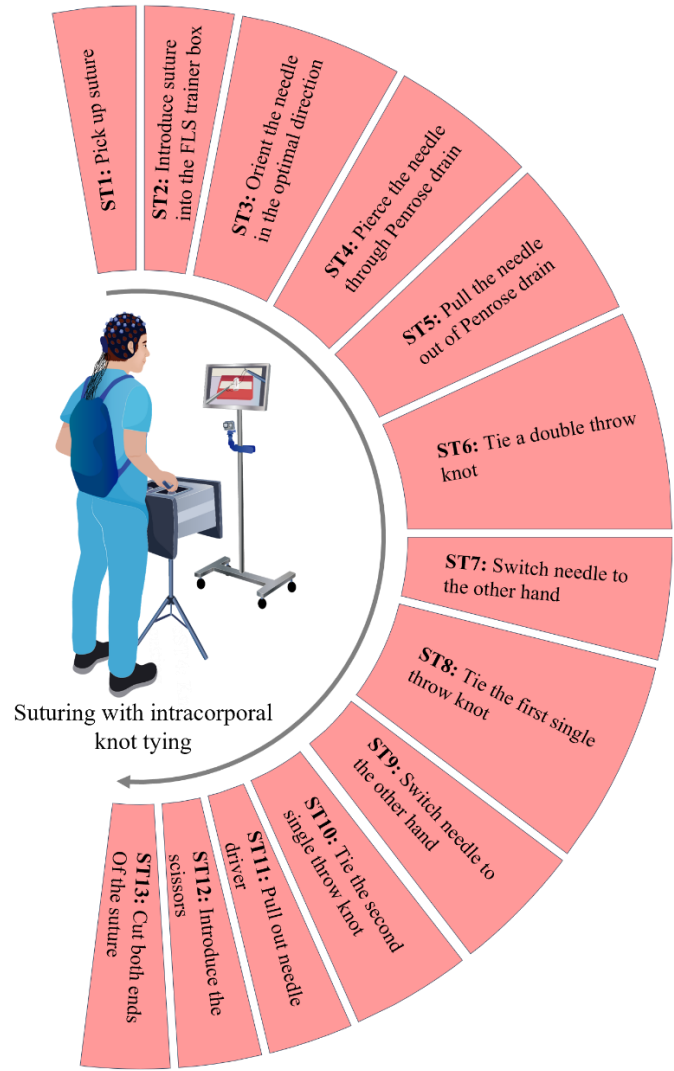

**Figure 3:** Hierarchical task analysis of the FLS suturing and intracorporal knot-tying task into thirteen subtasks (ST1-13)

## 2.6. Directed functional connectivity via non-linear Granger causality

To objectively quantify dFC between brain regions of interest, we utilized the non-linear Granger causality (nGC) framework developed in our previous work[20]. This model replaces the vector autoregressive (VAR) component of linear Granger causality[66][67] with an attention-based LSTM model[68][69], enabling it to capture long-term temporal dependencies between brain regions. For a detailed description of the method, refer to[20]. In this study, the nGC method was used to compute the inter-region brain connectivity at the subtask level across all the training and retention days.

### 2.7. 1D CNN with recursive feature elimination (RFE)

We used a 1D CNN classifier to evaluate the efficacy of the dFC in measuring motor skill learning (i.e. pre-training versus post-training), and retention (i.e. retention versus pre-and post-training) studies. It was further used to compare expert versus pre-and post-training. The architecture of the 1D CNN model and training process is explained in our previous study[20]. The input to the model is a set of twenty connectivities (dFC) obtained from the LSTM-based nGC model. To identify dominantly/highly adapted connectivities in motor skill learning and retention studies, we used the recursive feature elimination (RFE) approach[70][71] for each subtask separately explained in[20].

Rather than selecting an arbitrary threshold on the accuracy of classification to identify the learned subtasks, we used the accuracy of expert and novice surgeon classification for suturing and intracorporal knot-tying tasks from our previous study[20]. Experts and novices, being at the extremities of the learning curve, exhibit greater differences in neural interaction, allowing us to set a higher bar on the threshold. The expert and novice groups were differentiated with an accuracy of 82.8% which was used as the threshold in this study. This threshold is close to the threshold of 80% used in other clinical[72][73] and non-clinical[74] studies.

### 2.8. Change point detection

We applied the changepoint detection (CPD)[75][76] method to identify the specific day when substantial improvement in participants' neural process and performance occurred which marked the transitions in their learning stage of the Fitts and Posner model. Specifically, we detected the abrupt improvement in the means of the dFC and performance data (i.e. the FLS score and the task duration) over the course of training. The details of the CPD method can be found in Killick et al,[77]. Note that in the case of dFC, the CPD analysis was performed on the highly adapted connectivities for respective subtasks.

## 3. Results

### 3.1. Evaluation of motor skill learning

To quantify and evaluate the neural adaptation as a result of motor learning, we compared the dFC between pre- and post-training days at the subtask level. A high classification accuracy would indicate successful neural adaptation resulting from learning. A separate 1D-CNN binary classifier was trained for each subtask. The performance metrics of the trained classifiers are presented in Table 1. Notable neural adaptation was observed in subtasks ST3, ST4, ST6, ST8, and ST13, as highlighted in Table 1. All the performance metrics, including accuracy, sensitivity, specificity, Matthew's correlation coefficient (MCC), and area under the curve (AUC), were consistently high, indicating robust adaptation in dFC. The confusion matrix and receiver operating characteristic curves (ROC) for the classifiers are provided in the supplementary materials (S. Figure 1).

| Subtasks | Accuracy | Sensitivity | Specificity | MCC | AUC |
| --- | --- | --- | --- | --- | --- |
| ST1 | 0.740 | 0.893 | 0.414 | 0.354 | 0.644 |
| ST2 | 0.662 | 0.718 | 0.543 | 0.253 | 0.619 |
| ST3 | **0.859** | **0.892** | **0.776** | **0.658** | **0.867** |
| ST4 | **0.832** | **0.839** | **0.815** | **0.627** | **0.856** |
| ST5 | 0.749 | 0.845 | 0.537 | 0.396 | 0.733 |
| ST6 | **0.873** | **0.939** | **0.696** | **0.669** | **0.844** |
| ST7 | 0.787 | 0.796 | 0.767 | 0.530 | 0.824 |
| ST8 | **0.835** | **0.869** | **0.735** | **0.582** | **0.842** |
| ST9 | 0.800 | 0.865 | 0.612 | 0.477 | 0.778 |
| ST10 | 0.750 | 0.814 | 0.545 | 0.345 | 0.743 |
| ST11 | 0.707 | 0.797 | 0.419 | 0.211 | 0.581 |
| ST12 | 0.806 | 0.897 | 0.523 | 0.446 | 0.749 |
| ST13 | 0.944 | 0.978 | 0.841 | 0.846 | 0.957 |

**Table 1:** Comparison of dFCs between pre- and post-training across various subtasks.

Recursive Feature Elimination (RFE) identified specific dFCs that showed significant adaptation at the subtask level. This granular analysis revealed connectivity patterns that might not have been detected in task-level analysis. The dominantly adapted dFCs are shown in Figure 4. Notably, the LPFC → SMA, SMA →LPMC, and LPFC→RPFC were among the most highly adapted dFCs, as further discussed in the "Discussion" section.

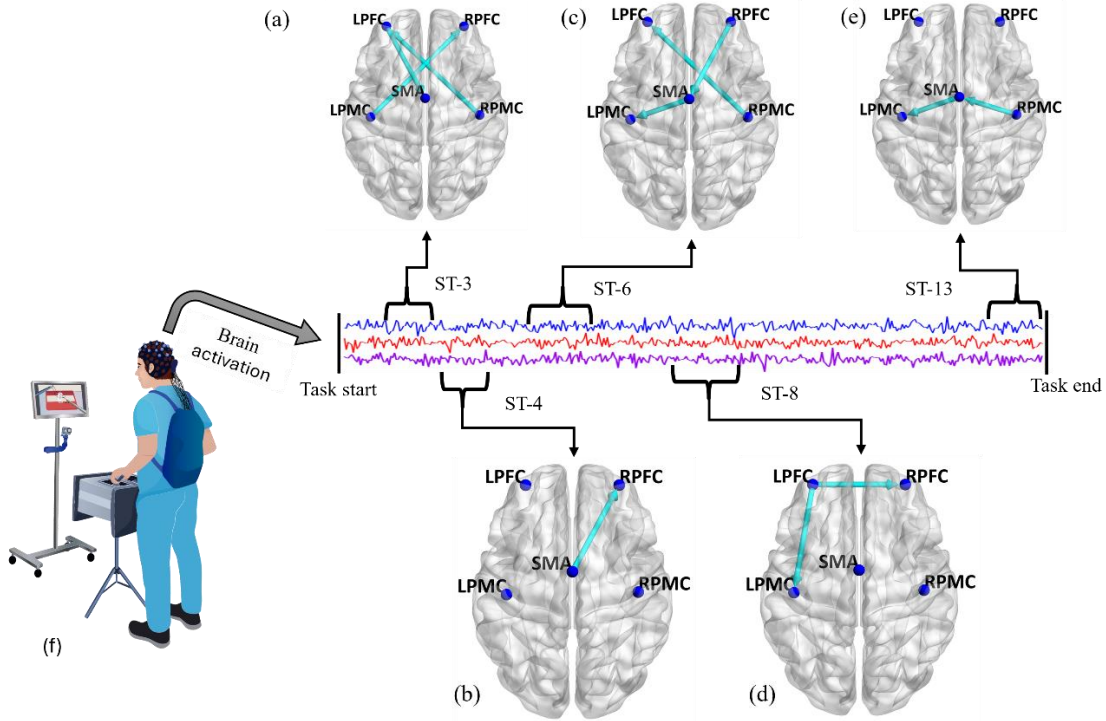

**Figure 4:** Dominantly adapted dFCs from pre- to post-training as identified by RFE for subtask (a) ST3, (b) ST4, (c) ST6, (d) ST8, and (e) ST13. Arrows indicate the source and target brain regions.

To assess the sensitivity of the dFC in measuring neural adaptation induced by motor learning, we compared the dFCs of the control group on pre-training and post-training days. In contrast to the learning group, the control group showed no significant adaptation in dFCs across any subtasks, as presented in S. Tables 4.

To further understand how the dFC evolved over consecutive days of training, we tracked changes in the strength of highly adapted dFCs for each subtask. This analysis allowed us to observe the gradual adjustments in dFC strength throughout the learning process. Notably, a steady decrease in the strength of dFC (LPFC→SMA) for subtask 3 was observed as training progressed from the pre- to the post-training day (Figure 6). To emphasize this trend, the mean values of dFC are presented by a connecting line across the training days. Similar decreases in other highly adapted dFCs for other subtasks were also observed and are provided in the supplementary materials (S. Figure 2). To examine neural adaptation at the individual level, we employed CUSUM plots[78] to identify trainees who crossed the "$h_0$" threshold and those who did not. The "$h_0$" represents a predefined decision boundary used to detect significant changes in the learning performance over time. A threshold of intermediate skill level (i.e. 280.8 [79]) on the FLS score was used to classify the performance/trial of trainees as unsuccessful or successful that was used in computing the CUSUM score. Most trainees demonstrated improvement with training at around 20 trials, as evidenced by a negative slope[79] in their CUSUM plots (S. Figure 3). To emphasize the inter-trainee variability in learning and neural adaptation at the individual level, we selected two trainees: P02, who crossed the "$h_0$" threshold, and P16, who did not. Figure 5 presents their CUSUM scores, and Figure 7 shows the evolution of dFC for P02 and P16 during ST8. The evolution of the remaining adapting connectivities for other subtasks is shown in S. Figure 4. Notably, P02 demonstrated a decreasing (adapting) trend in dFC across all subtasks, indicating successful neural adaptation throughout the training process. In contrast, P16 exhibited a non-decreasing trend in dFC for ST8, and ST13(Figure 7 and S. Figure 4(e)), suggesting less effective neural adaptation in these specific subtasks.

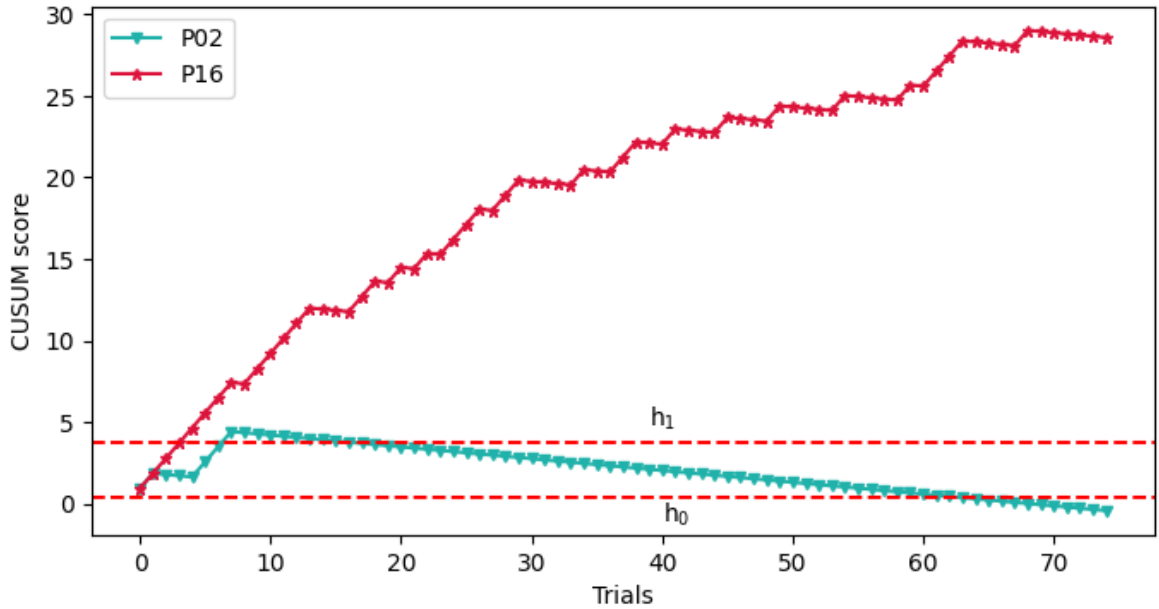

**Figure 5:** CUSUM scores of trainees P02 and P16. The horizontal dashed lines indicate the "$h_0$" and "$h_1$" thresholds.

### 3.2. Identification of motor learning stages

Experts, with years of experience, have mastered the skill and are in the autonomous stage of motor skill learning[80][81]. Trainees, on the other hand, are in the process of learning and progressing through various stages of the Fitts and Posner model. To evaluate whether the trainees have reached the autonomous stage, we compared brain connectivity from pre-training and post-training days to those of expert participants. The results are presented in the supplementary materials (S. Figure 5). The high classification accuracy and its decreasing trend going from pre-training to post-training (S. Figure 5) indicate that while the trainees are acquiring the skill and getting closer to the expert, they have not yet reached expert-level proficiency, and thus, they did not transition to the autonomous stage. Thus, we evaluated the potential of dFC for identifying the day of transition between the first two stages (i.e., cognitive and associative) by using the CPD method to the mean dFC data across training days. The analysis revealed that the transition from the cognitive to the associative stage occurred on the sixth day of training for several subtasks, including ST3 (Figure 6), ST4, ST6, and ST8 (S. Figure 2(c-h)) for most of the highly adapted dFCs. However, subtask ST13 (S. Figure 2(i-j)) showed an earlier transition, occurring on the third day of training. Further, the transition day is highlighted in the individual-level evolution of dFC in Figure 7 for the trainees P02 and P16.

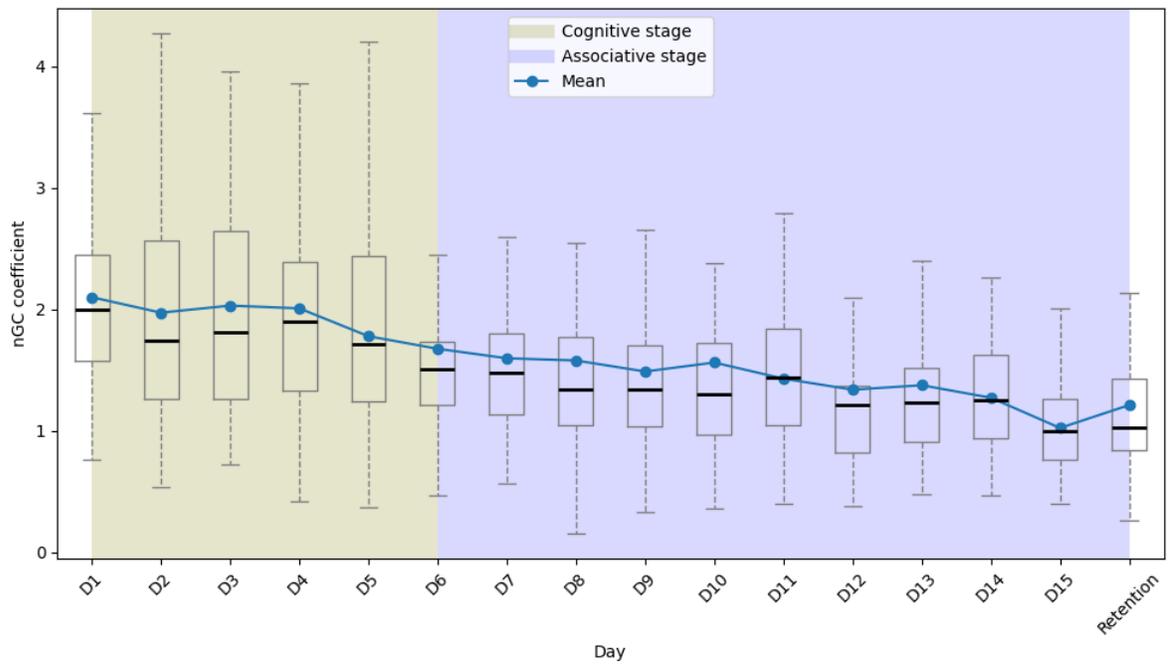

**Figure 6:** Distribution of dFC strength from LPFC→SMA for subtask 3 (ST3) on training and retention days. The mean dFC values across training days are connected by lines, illustrating a gradual decrease in its strength over time. The transparent background regions represent motor learning stages, with a transition from the cognitive to the associative stage observed on the sixth day of training based on dFC.

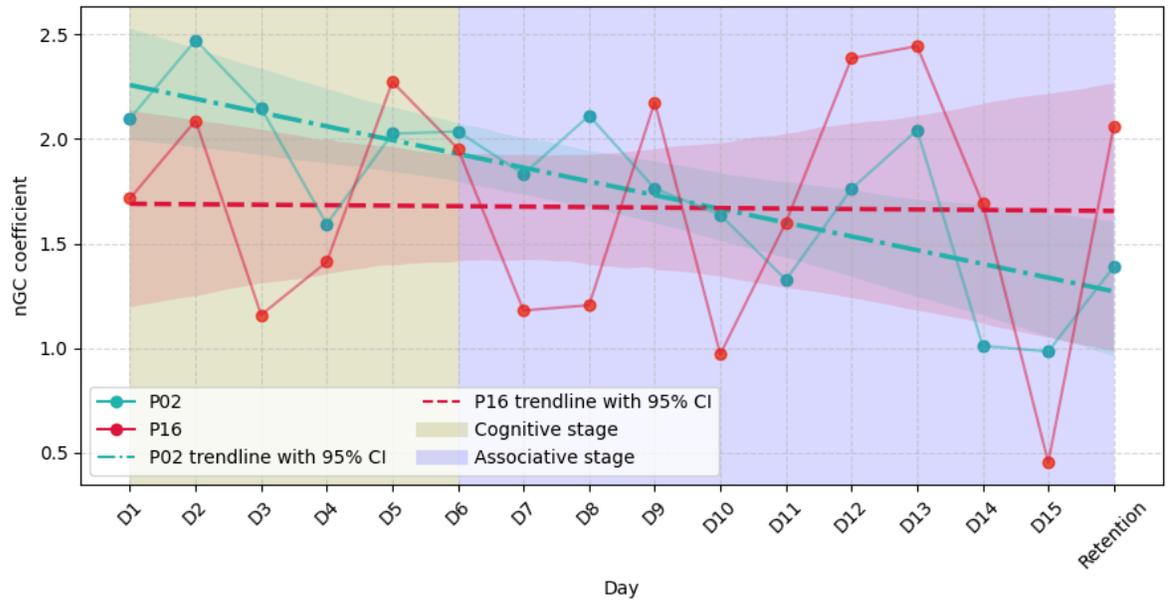

**Figure 7:** Evolution of dFC for trainees P02 and P16 for adapting subtasks ST8 for LPFC→LPMC. Linear regression is used to show the trend in the dFC over training days for both the trainees, with 95% confidence interval (CI) shown in shaded regions around the trend line. The transparent background regions represent motor learning stages, with a transition from the cognitive to the associative stage observed on the sixth day of training based on dFC.

We also applied the CPD method to the performance score (FLS score) to identify the cognitive and associative stages. The CPD analysis identified the fifth day as the transitioning day where the participants progressed from the cognitive stage to the associative stage of motor learning (Figure 8(a)). We further applied CPD to the task duration, which also indicated the fifth day as the transition point (Figure 8(b)). Both behavioral metrics consistently indicated that participants transitioned to the associative stage of motor learning on the fifth day. This transition is characterized by a rapid increase in FLS score and a decrease in task duration prior to it, followed by slower improvements thereafter.

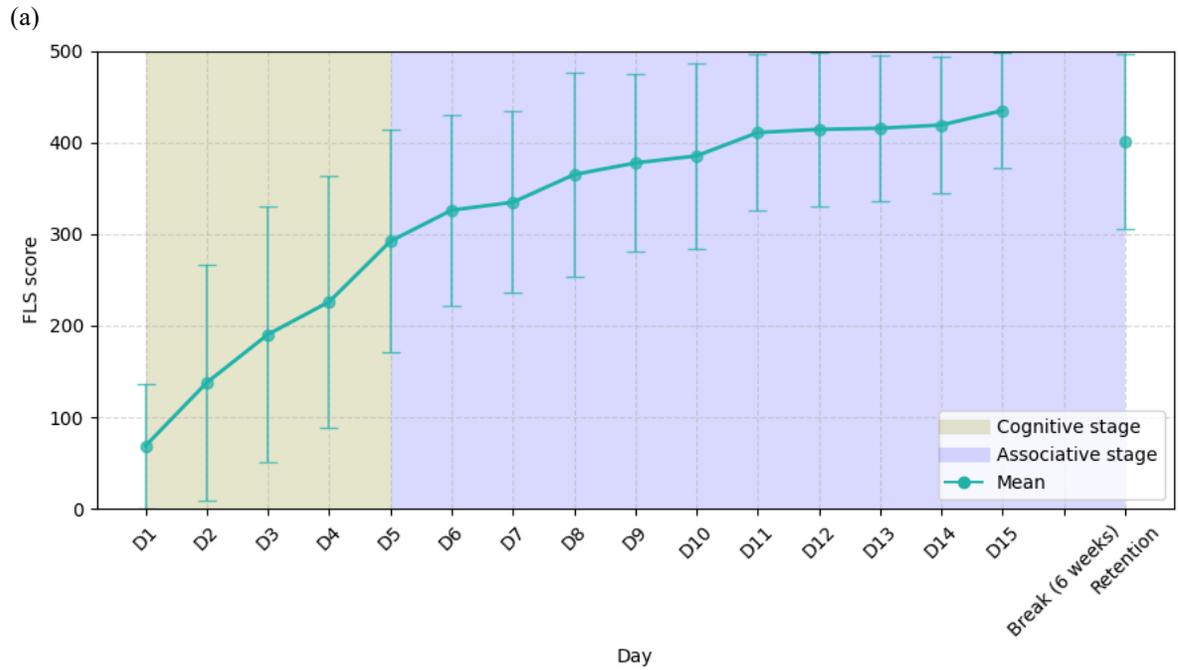

(a)

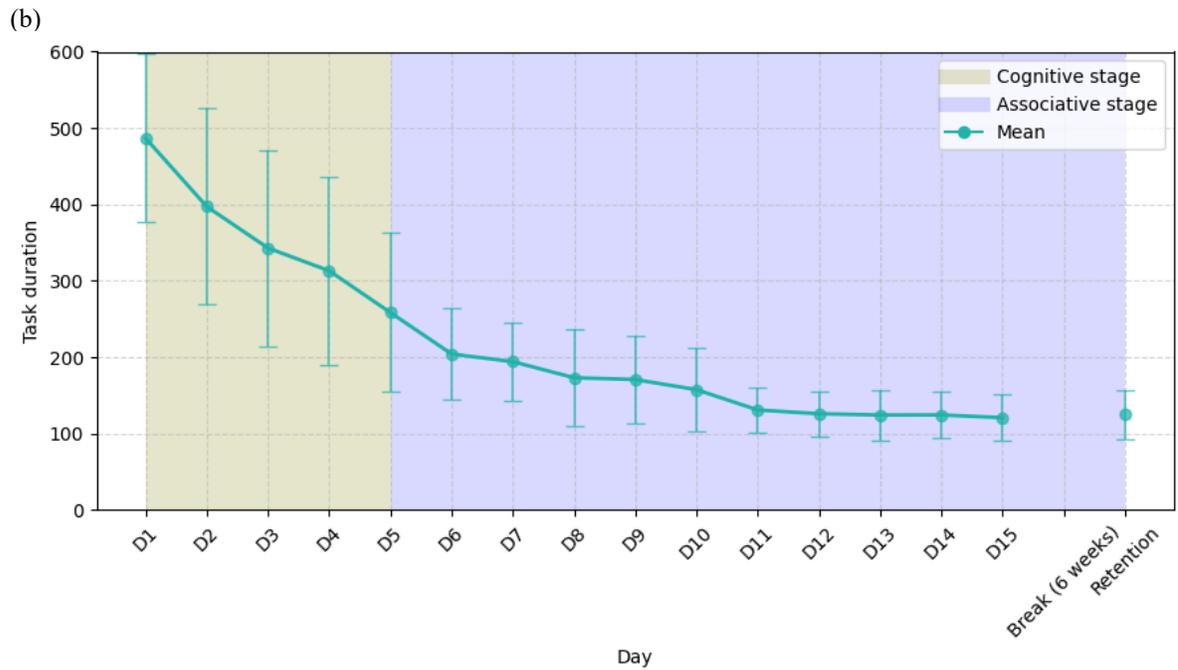

(b)

**Figure 8:** Mean of (a) FLS score and (b) task completion time of the trainees over training days. The transparent background regions represent motor learning stages, with a transition from the cognitive to the associative stage observed on the fifth day of training. The bar indicates the standard deviations in performance.

Note that both the dFC and performance-based CPD analysis identified trainees to be in cognitive and associative stages on the pre-training and post-training days, respectively and unable to reach the autonomous stage.

### 3.3. Evaluation of motor skill retention

In addition to evaluating motor skill learning, we investigated whether dFC can reveal the retention of newly acquired motor skills. To assess this, we examined the stability of the dFCs established at the end of training over a six-week washout period. Specifically, we compared the dFCs on the retention day with the pre-training and the post-training days.

No significant differences were observed between the dFCs on the retention day and post-training day, indicating that the neural adaptations achieved at peak performance were maintained over the six-week washout period (see supplementary materials S. Table 5). On the other hand, significant differences were found between dFCs on pre-training (i.e. prior to consolidation) and retention day for subtasks, as highlighted in Table 2. The confusion matrix and the ROC curve for the discriminating dFCs are provided in supplementary materials (S. Figure 6). Furthermore, the strength of dFC on the retention day was similar to that of the post-training day, as shown in Figure 6 (see retention day and day 15, two-sample t-test, two-tailed, $p > 0.01$, $t = 0.234$, $df1 = 148$, $df2 = 61$), confirming the stability of functional connectivity associated with skill retention.

| Subtasks | Accuracy | Sensitivity | Specificity | MCC | AUC |
|---|---|---|---|---|---|
| ST1 | 0.667 | 0.806 | 0.543 | 0.359 | 0.665 |
| ST2 | 0.606 | 0.677 | 0.543 | 0.222 | 0.619 |
| ST3 | **0.875** | **0.871** | **0.879** | **0.750** | **0.896** |
| ST4 | 0.810 | 0.787 | 0.831 | 0.619 | 0.841 |
| ST5 | 0.814 | 0.855 | 0.776 | 0.631 | 0.849 |
| ST6 | **0.862** | **0.933** | **0.786** | **0.730** | **0.889** |
| ST7 | 0.808 | 0.750 | 0.867 | 0.621 | 0.852 |
| ST8 | **0.844** | **0.867** | **0.816** | **0.684** | **0.863** |
| ST9 | 0.787 | 0.949 | 0.592 | 0.590 | 0.764 |
| ST10 | 0.748 | 0.746 | 0.750 | 0.492 | 0.763 |
| ST11 | 0.627 | 0.492 | 0.814 | 0.314 | 0.606 |
| ST12 | 0.743 | 0.947 | 0.477 | 0.495 | 0.743 |
| ST13 | **0.830** | **0.857** | **0.795** | **0.654** | **0.845** |

**Table 2:** Comparison of dFCs between pre-training and retention day across various subtasks.

Using RFE, we identified the highly adaptive (i.e. dominantly differentiating) dFCs between the pre-training and retention days at the subtask level. These adaptive connectivities are illustrated in Figure 9.

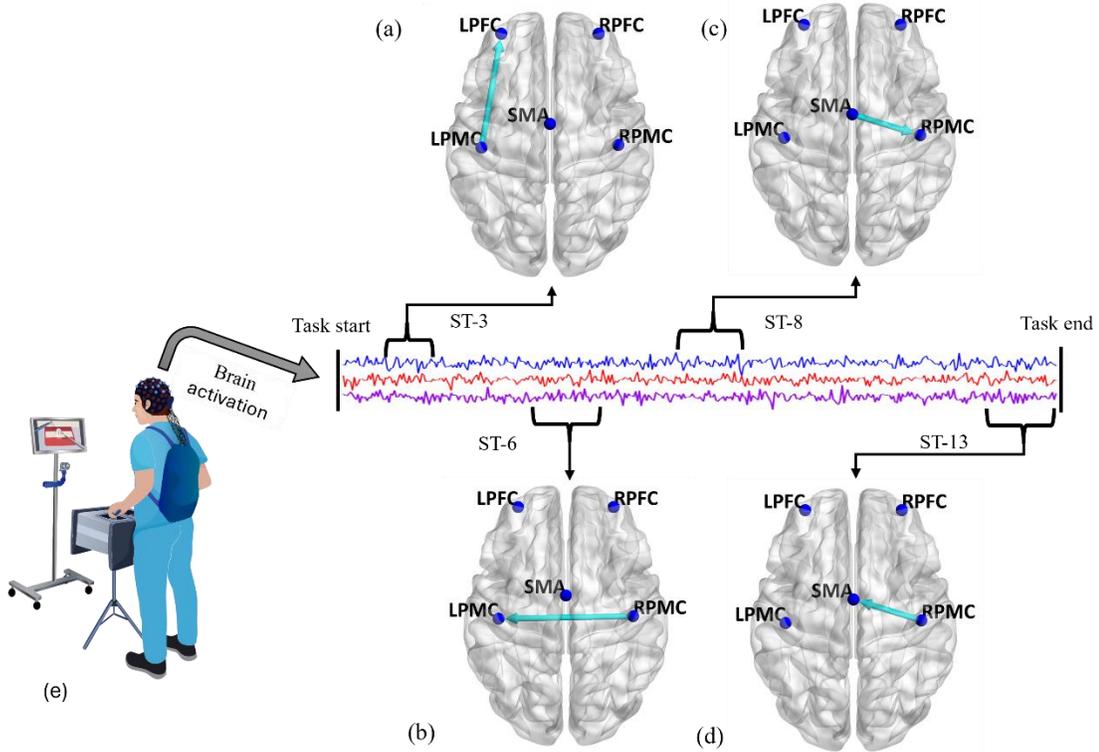

**Figure 9:** Dominantly differentiating dFCs from pre-training to retention day, identified using the RFE method for ST3(a), ST6(b), ST8(c), and ST13(d).

### 3.4. Evaluation of motor skill learning with FLS score

To assess the utility of the FLS performance score in capturing motor skill learning and retention, a 1D-CNN classifier was trained using the FLS scores. Although the FLS score aggregates performance across the entire trial and may lack specificity in evaluating individual subtasks, it remains a standard measure for overall skill assessment. The results of this analysis are presented in Table 3.

|  | **Accuracy** | **Sensitivity** | **Specificity** | **MCC** | **AUC** |
| --- | --- | --- | --- | --- | --- |
| Pre-training versus post-training | 0.845 | 0.718 | 0.981 | 0.719 | 0.936 |
| Pre-training versus Retention | 0.885 | 0.867 | 0.893 | 0.743 | 0.916 |
| Post-training versus Retention | 0.604 | 0.480 | 0.659 | 0.132 | 0.554 |

**Table 3:** Performance metrics of motor skill learning, and retention based on FLS score.

Statistical analysis revealed that FLS scores on the pre-training day were significantly lower than both the post-training (two-sample t-test, two-tailed, $p < 0.01$, $t = -35.91$, $df1 = 169$, $df2 = 158$) and the retention days (two-sample t-test, two-tailed, $p < 0.01$, $t = -21.33$, $df1 = 158$, $df2 = 74$), indicating an improvement in skill proficiency due to training. No significant difference was found between the post-training and the retention day scores (two-sample t-test, two-tailed, $p > 0.01$, $t = 2.53$, $df1 = 169$, $df2 = 74$), suggesting effective retention of motor skills over the washout period.

## 4. Discussion

The findings of this study highlight the utility of directed functional connectivity (dFC) as a robust biomarker for understanding the neural mechanisms underlying motor skill learning and retention. By capturing the dynamic neural adaptations that occur during the acquisition of complex motor skills, dFC provides insights beyond traditional performance metrics, offering a comprehensive view of the learning process. The results revealed distinct neural patterns associated with different stages of the Fitts and Posner model, underscoring the progressive refinement of motor control circuits as participants transitioned from the cognitive to the associative stage. Additionally, the study demonstrated the stability of dFC patterns over an extended washout period, suggesting that neural adaptations associated with skill learning are well-retained. These findings have important implications for designing personalized, brain-based training protocols to optimize motor skill acquisition and retention, particularly in fields requiring high precision, such as surgical education. Below, we discuss the significance of these results, their alignment with existing literature, and the broader implications for skill acquisition research.

The large difference in post-training and expert brain connectivity (S. Figure 5) indicates that trainees did not reach the autonomous stage. This outcome can be attributed to two key factors that likely influenced the depth of skill acquisition. First, the complexity of the suturing with intracorporeal knot-tying task likely demands prolonged and intensive training to achieve the autonomous stage. Unlike simpler motor tasks that may lead to automation within hours or weeks of practice, mastering complex surgical skills often requires months of dedicated effort[82][83]. Second, the observed differences in brain connectivity suggest that the neural adaptations associated with skill automation had not fully developed within the training period of this study. Neural adaptation involves refining motor control circuits and reducing reliance on higher-order cognitive processes, which typically occur over extended practice. Together, these findings underscore the need for longer training durations and more repetitions to achieve the autonomous stage in complex tasks like this, where precision and coordination are paramount.

The dFC-based CPD analysis identified the transition from the cognitive to the associative stage on the sixth day of training for most of the subtasks, while behavioral metrics indicated this transition on the fifth day. This temporal discrepancy suggests a potential lag between observable performance improvements and underlying neuroplastic changes. The delay may be attributed to motor memory consolidation during sleep or rest periods following the training on the fifth day. Consolidation not only cements neural processes but also enhances them[28]. The "one-day" delay aligns with the time typically needed for sleep-dependent consolidation processes to occur[28]. This finding aligns with other studies that have observed delayed neuroplasticity or neural processes following motor skill training[35][84][85], which is speculated to relate to the fine-tuning of movement patterns, leading to reduced energy and metabolic costs[86]. The variation in transition days between subtasks (sixth day for most and third day for ST13, S. Figure 2) suggests that motor learning unfolds over different timescales for different subtasks and dFCs[87][22]. The dFC results can be used in conjunction with behavioral metrics to provide a more comprehensive understanding of skill acquisition. The early transition on the third day for subtask ST13 provided extended time for neural adaptation in the associative stage, making it the most differentiating subtask between pre- and post-training days, with a high classification accuracy of 94.4%. The initial rapid improvement of performance in the cognitive stage, followed by a gradual gain in the associative stage (Figure 8 (a) and (b)) aligns with the typical learning curve reported by other studies that are consistent with the Fitts and Posner model[23][88][89]. Early gains likely result from initial task comprehension and strategy formation, while later gains reflect refined motor control through practice. Both online[90] and offline[91][92](i.e. the resting period between trials) learning may have contributed to the observed continuous improvement.

The high-performance metrics (accuracy, sensitivity, specificity, MCC, and AUC) achieved by the classifiers (Table 1) in distinguishing pre- and post-training dFC patterns underscore the efficacy of dFC as a biomarker of motor skill learning. The observed difference is due to the neural adaptation induced by motor skill learning on progressing from the start of the cognitive to the end of the associative stage[16][24][93]. High neural adaptation was observed during specific subtasks (ST3, ST4, ST6, ST8, and ST13), likely due to their complexity, requiring greater cognitive and motor adjustments compared to simpler subtasks. These findings are consistent with previous studies[24][94][95] that reported neural adaptations associated with the acquisition of new motor skills. The lack of neural adaptation in the control group (S. Table 4) confirms that the dFC changes in the learning group were specifically induced by motor training, rather than external factors such as task familiarity or incidental exposure. This supports the specificity of dFC as a sensitive and reliable biomarker of motor learning.

The gradual decrease in strength of highly adapted dFCs from pre- to post-training (Figure 6 and S. Figure 2) is consistent with Hebb's law[96], which posits that that repeated activation of neurons during training leads to more efficient communication[97], reducing the overall strength of information flow[98]. This observation aligns with other studies[99][100][97][101][102][103][35] that observed a decrease in connectivity strength as a result of neural adaptation[104], which increases efficiency in neural circuits[102][101][105]. For instance, Deeny et al reported a decrease in alpha-band coherence between motor planning and visuospatial attention regions in highly skilled performers[106]. Moreover, motor training reduces the overall excitability threshold of the motor cortex[107], thus requiring less information from the source brain region. Yet it should be noted that the increase or decrease in strength of connectivity is highly task-specific and associated cognitive demands[108]. The correlation between dFC adaptations and behavioral improvements in performance scores (Figure 8(a) and(b)) indicates that dFC is directly associated with performance gains. The adaptation likely lead to the observed reduction in the variability in task duration which is considered the hallmark of skill learning[90]. Further, we observed that group-level neural adaptation is reflected at the individual level (Figure 7 and S. Figure 4), making it a reliable biomarker for tracking individual progress. For instance, if connectivity remains high, it may indicate that the trainee is struggling or not consolidating skills efficiently, providing a quantifiable biomarker for identifying when intervention is necessary. This insight can be leveraged to design personalized training protocols with targeted interventions to enhance the training. Thus, dFC can serve as a biomarker for real-time feedback, enabling trainers to adjust the intensity, duration, and brain stimulation[109][110], or focus on specific subtasks based on individual neural adaptation patterns. Such tailored interventions ensure that each trainee receives customized support to optimize their learning trajectory.

The HTA identified key subtasks at which the adaptation of dFC occurred. For instance, in subtask 3, LPFC → SMA, LPMC → RPFC, and RPMC → LPFC were the highly adapted dFCs from pre- to post-training (Figure 4(a)). The presence of multiple dFCs suggests the integration of various neural mechanisms that collectively contribute to motor skill learning[111]. In this subtask, the participants need to orient the needle (held with one hand) perpendicular to the Penrose drain (held with the other hand). The role of LPFC is to generate motor plans, and cognitive control for motor outputs which are transferred to the SMA, responsible for the generation and coordination of sequential motor tasks[112] and their timing for aligning the needle perpendicular to the Penrose drain. Research has shown that motor learning involves the LPFC → SMA, particularly during the early stages of learning, where higher-order cognitive processes are needed to organize movements into coherent sequences[1][103]. The observed adaptation is likely due to the difference in top-down control of LPFC in the cognitive and associative stages[113]. LPMC and RPMC receive sensory feedback from effectors[114][115] regarding the current position of the needle which is transferred to the prefrontal cortex (RPFC and LPFC) to update the cognitive plan and to orient the needle to be perpendicular to the Penrose drain while holding the Penrose drain with another hand[116]. Here, the sensory feedback helps in real-time adjustment[117]. These adapted connectivities likely reflect an improvement in the integration of sensory feedback (from the RPMC and LPMC) and cognitive control processes (to LPFC and RPFC), enabling efficient control of the needle's orientation with increasing experience [118]. This observation aligns with the dorsal stream theory of visuomotor control[103][119], where feedback from the motor regions is processed by the prefrontal cortex to guide adjustments based on somatic feedback. Overall, the observed adaptation aligns with the participants progressing through different stages of motor learning[23], with the brain regions being recruited differently[120].

In subtask 4, while piercing the needle into the Penrose drain, SMA → RPFC (Figure 4(b)) was the highly adapted connectivity between the cognitive and associative stages. The SMA, involved in the timing and coordination of fine motor movements, sends feedback to the RPFC[121], which plays a key role in spatial attention (to locate and focus on the marks on the Penrose drain) and monitoring the accuracy of the needle's trajectory and timed application of the force required to pierce the stretched drain[122][123]. In subtask 6, while tying the knot, one of the highly adapted connectivities, i.e. RPFC → SMA (Figure 4(c)) transfers the attention and strategic plan to the SMA for coordination of motor actions[121]. The second connectivity i.e. SMA → LPMC links cognition to movement[124][125]. It facilitates the transfer of motor sequences to the motor region[126] for execution while tying the knot. Subsequently, the LPMC directs specific motor commands, such as motor angle and required muscle forces, to the hands via the corticospinal tract, to ensure accurate execution of the movement. The third connectivity RPMC → LPFC reflects the difference in sensory feedback dominantly from the hands to the LPFC for planning and cognitive control[127]. In subtask 8, while tying a single throw knot involved two adaptive dFCs: LPFC → LPMC and LPFC → RPFC (Figure 4(d)). The connectivity LPFC → LPMC is essential to transferrin motor plans to the motor region[128][129] for the precise execution required in manipulating the thread with precision to form the knot. Simultaneously, the connectivity LPFC → RPFC transfers the motor plans to the RPFC for spatial attention[130] to crucial environment cues to carefully manage the orientation and positioning of the thread while tying the knot. The RPMC → SMA and SMA → LPMC (Figure 4(e)) were the adaptive connectivities in subtask 13, which involves cutting both ends of the suture. RPMC receives sensory inputs from the somatosensory region[131] and transfers them to the SMA for coordinating, timing, and sequencing the motor actions for precise motor control and the overall coordination of cutting. Simultaneously, the SMA's communication with the LPMC underscores the need for precise control and timing of right-hand movements, which are crucial for properly orienting the scissors, holding the suture and executing the cut. This connectivity aligns with the SMA's known role in coordinating complex bimanual tasks[126].

The absence of differences in dFC for any subtask between the last day of the associative stage (i.e. post-training) and the retention day (S. Table 5) suggests that neural adaptations are maintained over the six-week washout period (see strength of connectivities on post-training and retention days in Figure 6 and S. Figure 2, and the statistical test ($p > 0.01$) in "*Evaluation of motor skill retention*" section), indicating successful consolidation. This stability is supported by consistent performance scores between post-training and retention days (Table 3, row 3 and the statistical test ($p > 0.01$) in "*Evaluation of motor skill learning with FLS score*" section), underscoring dFC's potential as a reliable biomarker for long-term motor skill retention. The observed retention may be attributed to the successful consolidation of the adapted neural interactions[132][133][134]. These findings suggest that once learners transition from the cognitive to the associative stage, the learned motor skills become ingrained at a neural level, allowing for sustained performance without substantial skill decay[28]. However, clear differences in dFCs were observed between pre-training and retention (Table 2) for several subtasks. The differentiating subtasks were similar to that of the learning (i.e. pre- versus post-training) analysis (see Table 1 and Table 2, except ST 4), further supporting the similarity in dFC between post-training and retention day and consequently the stability of the adapted dFCs from the end of training to the retention day. The difference is reflected in the performance score (Table 3, row 2) which also shows clear differences in pre-training and retention. Interleaved training in this study may have facilitated successful retention, as corroborated by previous studies [36]. Research on motor skill consolidation has consistently shown that rest periods or sleep significantly enhance the retention of newly acquired motor skills[135]. Sleep-dependent mechanisms, such as spindle oscillations[136][137], may have played a key role in this process by facilitating neural consolidation and stabilization of the learned skill. These findings emphasize the importance of rest following active practice in facilitating consolidation which is important for long-term motor skill retention.

Analysis of retained dFCs further identified key connectivities supporting long-term skill retention. For instance, in subtask 3 (orienting the needle), the adaptation of LPMC → LPFC connectivity (Figure 9(a)) supports the consolidation of motor skill learning[129][104]. Various studies have reported that the PMC is involved in the storage of motor memory[138] and retention of learned movements [139][140][141], which is transferred to LPFC for cognitive control [57][142] during the orientation task. Similarly, the bimanual coordination observed in tying a double throw knot in subtask 6 is facilitated by RPMC→LPMC connectivity (Figure 9(b))[143], suggesting that interhemispheric communication[144], mediated by the corpus callosum[145][146], supports motor skill retention. In subtask 8 while tying a single throw knot, SMA → RPMC (Figure 9(c)) was the highly adapted connectivity, and in subtask 13 while cutting both ends of the suture, RPMC → SMA (Figure 9(d)) was the highly adaptive connectivity. Tying a knot and cutting the suture involves a sequence of precisely coordinated hand movements. These findings align with the established roles of the SMA and PMC: SMA is essential for timing, initiation, and motor sequencing[147][148][149], while PMC plays a key role in the acquisition, storage, and retrieval of motor memory[117][150][131] necessary for the precise movements.

While this study demonstrates dFC's potential as a biomarker for motor skill learning and retention, some limitations warrant consideration. First, the relatively small sample size limits the generalizability of the findings, and future studies should include larger and more diverse populations to strengthen the validity and applicability of the results. Additionally, while the training duration encompassed key learning stages, a longer training period and extended follow-up are necessary to understand the complete motor skill acquisition process, stages and to evaluate retention beyond the initial six-week period. Another limitation is the specificity of the brain region selection; future studies should incorporate a more detailed selection of brain regions, including subcortical areas such as the cerebellum, striatum, and basal ganglia, which are known to play significant roles in motor coordination, procedural learning, and retention[120][151].

## 6. Supplementary materials

| Participant | Age | Gender | Dominant Hand | Med Student? | Year in Med School/ Residency | Experience with laparoscopic tools? | Year of experience lap | FLS experience | Years of FLS/FRS |
|---|---|---|---|---|---|---|---|---|---|
| C01 | 26 | M | R | yes | M2 | yes | <1 | yes | <1 |
| C02 | 18 | F | R | no | no | no | no | no | no |

| Participant | Age | Gender | Dominant Hand | Med Student? | Year in Med School/Residency | Experience with laparoscopic tools? | Year of experience lap | FLS/FRS experience | Years of FLS/FRS |
|---|---|---|---|---|---|---|---|---|---|
| C03 | 21 | M | R | yes | M1 | no | no | no | no |
| C04 | 19 | F | R | no | no | no | no | no | no |
| C05 | 21 | M | R | no | no | no | no | no | no |
| C06 | 20 | F | R | no | no | no | no | no | no |
| C07 | 24 | F | R | yes | M2 | no | no | no | no |
| C08 | 22 | F | L | no | no | no | no | no | no |
| C09 | 25 | M | R | no | no | no | no | no | no |
| C10 | 20 | F | R | no | no | no | no | no | no |
| C11 | 28 | M | L | no | no | no | no | no | no |
| C12 | 26 | F | R | no | no | no | no | no | no |
| C13 | 21 | M | R | no | no | no | no | no | no |
| C14 | 22 | F | R | no | no | no | no | no | no |
| C15 | 20 | F | L | no | no | no | no | no | no |
| C16 | 21 | F | R | no | no | no | no | no | no |
| C17 | 21 | F | R | no | no | no | no | no | no |
| C18 | 20 | M | R | no | no | no | no | no | no |
| C19 | 21 | F | R | no | no | no | no | no | no |
| C20 | 18 | F | R | no | no | no | no | no | no |
| C21 | 18 | M | R | no | no | no | no | no | no |
| C22 | 19 | M | R | no | no | no | no | no | no |
| C23 | 22 | F | R | no | no | no | no | no | no |
| C24 | 20 | F | R | no | no | no | no | no | no |
| C25 | 23 | F | R | no | no | no | no | no | no |
| C26 | 22 | M | R | yes | M1 | no | no | no | no |
| C27 | 23 | F | R | no | no | no | no | no | no |

**S. Table 1:** Demography of control participants.

| Participant | Age | Gender | Dominant Hand | Med Student? | Year in Med School/Residency | Experience with laparoscopic tools? | Year of experience lap | FLS/FRS experience | Years of FLS/FRS |
|---|---|---|---|---|---|---|---|---|---|
| P01 | 22 | F | R | yes | M2 | no | no | no | no |
| P02 | 24 | M | R | yes | M2 | no | no | no | no |
| P03 | 24 | F | R | yes | M1 | no | no | no | no |
| P04 | 23 | M | R | yes | M2 | no | no | no | no |
| P05 | 33 | F | R | no | no | no | no | no | no |
| P06 | 21 | F | R | no | no | no | no | no | no |
| P07 | 24 | F | R | no | no | no | no | no | no |
| P08 | 21 | M | R | no | no | no | no | no | no |
| P09 | 24 | F | R | yes | M1 | no | no | no | no |
| P10 | 23 | F | R | no | no | no | no | no | no |
| P11 | 20 | F | R | no | no | no | no | no | no |
| P12 | 20 | F | R | no | no | no | no | no | no |
| P14 | 19 | M | L | no | no | no | no | no | no |
| P13 | 18 | M | R | no | no | no | no | no | no |
| P15 | 27 | F | R | yes | M1 | no | no | no | no |
| P17 | 36 | M | R | no | no | no | no | no | no |
| P18 | 24 | F | R | yes | M1 | no | no | no | no |
| P16 | 27 | M | R | no | no | no | no | no | no |
| P19 | 22 | F | R | no | no | no | no | no | no |
| P20 | 25 | M | R | yes | M1 | no | no | no | no |
| P21 | 18 | F | R | no | no | no | no | no | no |
| P22 | 31 | M | L | no | no | no | no | no | no |
| P23 | 29 | M | R | no | no | no | no | no | no |
| P24 | 21 | F | R | no | no | no | no | no | no |
| P25 | 22 | M | R | yes | M1 | no | no | no | no |
| P26 | 20 | F | R | no | no | no | no | no | no |
| P27 | 22 | F | R | no | no | no | no | no | no |
| P28 | 20 | F | R | no | no | no | no | no | no |

**S. Table 2:** Demography of trainee participants.

| S.N. | EEG-directed functional connectivity |
|---|---|
| 1 | LPFC-->RPFC |
| 2 | LPFC-->LPMC |
| 3 | LPFC-->RPMC |
| 4 | LPFC-->SMA |
| 5 | RPFC-->LPFC |
| 6 | RPFC-->LPMC |
| 7 | RPFC-->RPMC |
| 8 | RPFC-->SMA |
| 9 | LPMC-->LPFC |
| 10 | LPMC-->RPFC |
| 11 | LPMC-->RPMC |
| 12 | LPMC-->SMA |
| 13 | RPMC-->LPFC |
| 14 | RPMC-->RPFC |
| 15 | RPMC-->LPMC |
| 16 | RPMC-->SMA |
| 17 | SMA-->LPFC |
| 18 | SMA-->RPFC |
| 19 | SMA-->LPMC |
| 20 | SMA-->RPMC |

**S. Table 3:** List of connectivities.

(a)
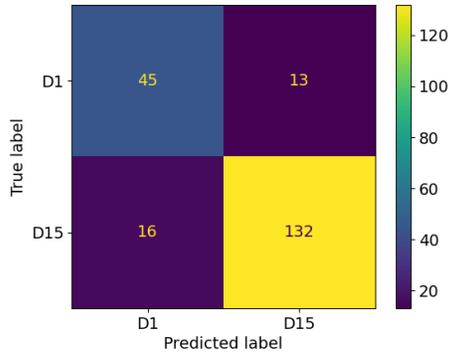

(b)
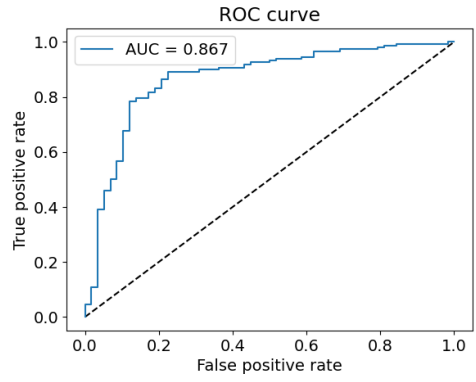

(c)
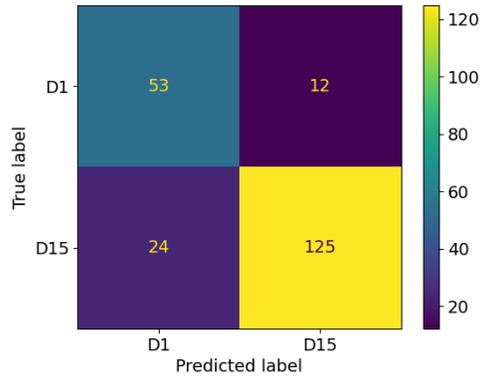

(d)
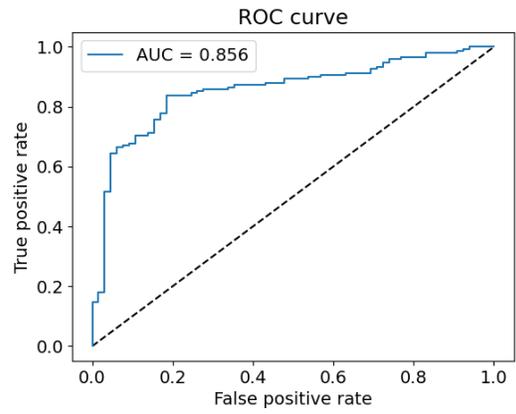

(e)
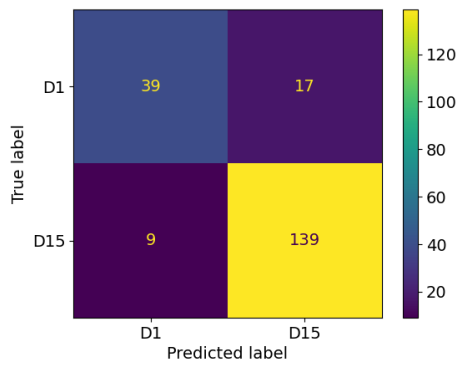

(f)
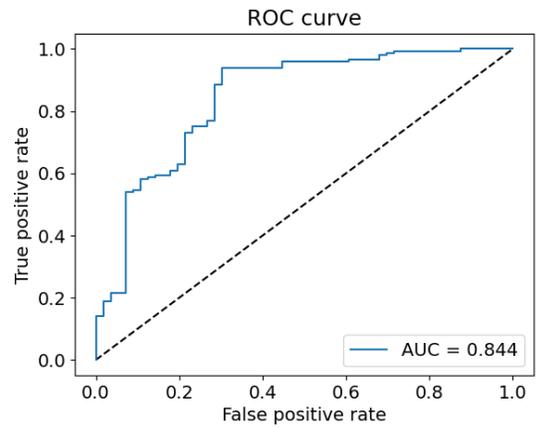

(g)

(h)

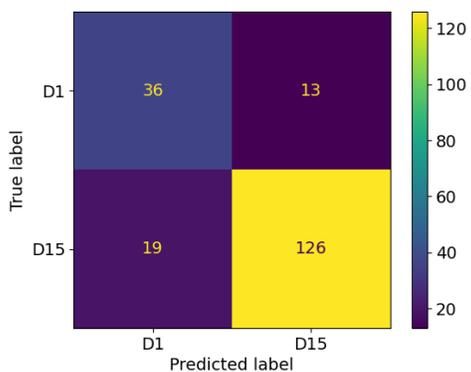
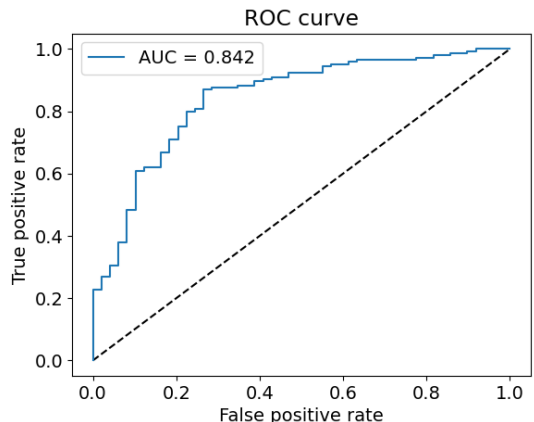

(i) (j)

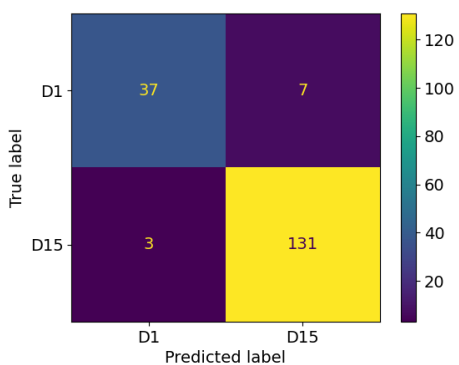
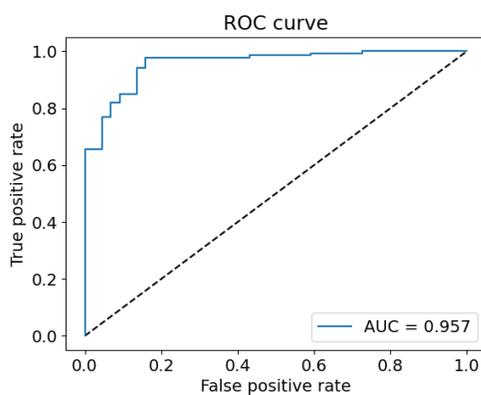

**S. Figure 1:** Confusion matrix and ROC curve in pre- versus post-training for (a-b) ST3 (c-d) ST4 (e-f) ST6 (g-h) ST8 (i-j) ST13.

| Subtasks | Accuracy | Sensitivity | Specificity | MCC | AUC |
| --- | --- | --- | --- | --- | --- |
| ST1 | 0.593 | 0.529 | 0.645 | 0.175 | 0.533 |
| ST2 | 0.632 | 0.373 | 0.841 | 0.244 | 0.573 |
| ST3 | 0.643 | 0.762 | 0.554 | 0.316 | 0.646 |
| ST4 | 0.688 | 0.604 | 0.754 | 0.362 | 0.682 |
| ST5 | 0.740 | 0.750 | 0.729 | 0.479 | 0.783 |
| ST6 | 0.625 | 0.452 | 0.783 | 0.250 | 0.598 |
| ST7 | 0.679 | 0.762 | 0.583 | 0.352 | 0.697 |
| ST8 | 0.653 | 0.559 | 0.737 | 0.301 | 0.669 |
| ST9 | 0.714 | 0.656 | 0.763 | 0.422 | 0.738 |
| ST10 | 0.735 | 0.969 | 0.528 | 0.544 | 0.797 |
| ST11 | 0.657 | 0.645 | 0.667 | 0.311 | 0.684 |
| ST12 | 0.733 | 0.679 | 0.781 | 0.463 | 0.756 |
| ST13 | 0.700 | 0.607 | 0.781 | 0.396 | 0.712 |

**S. Table 4:** Comparison of dFC of control group in pre-and post-training.

(a)

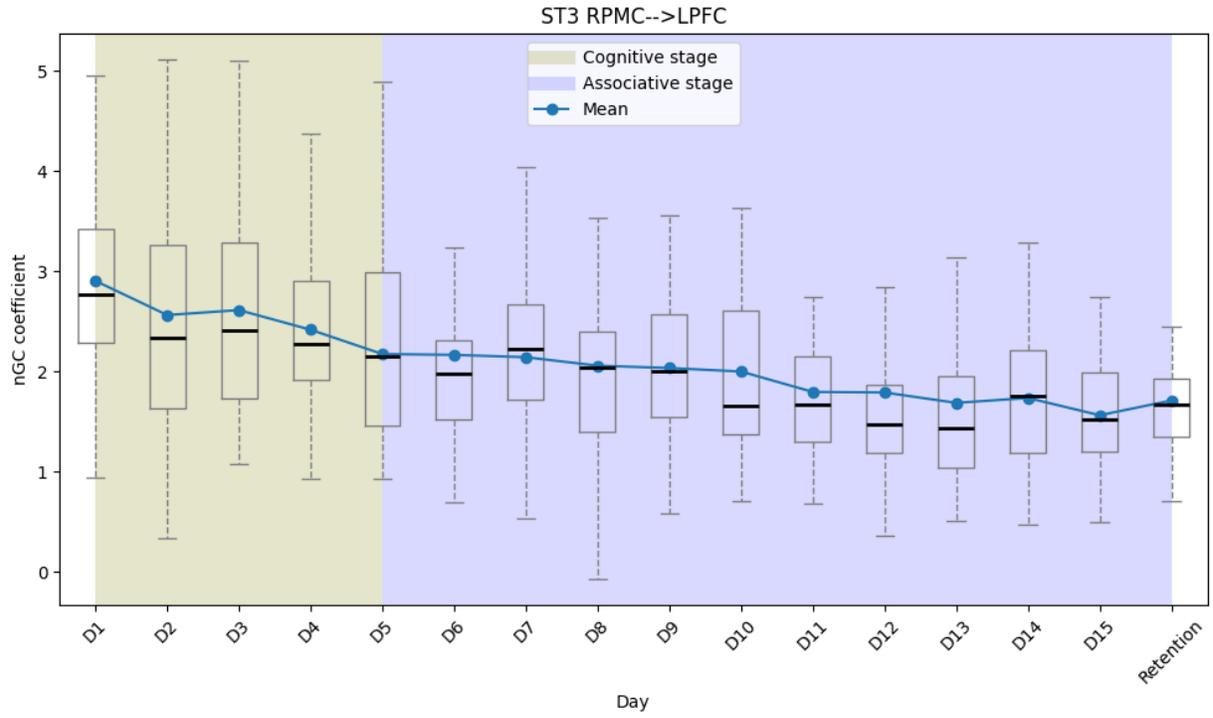

(b)

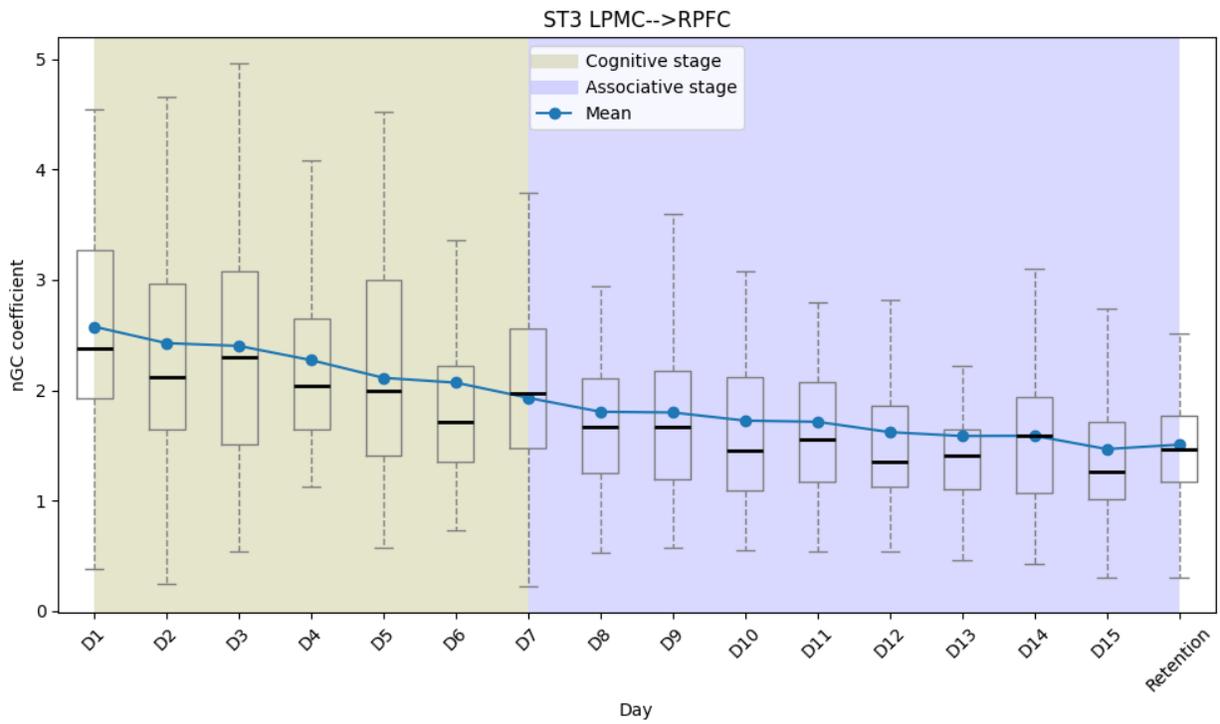

(c)

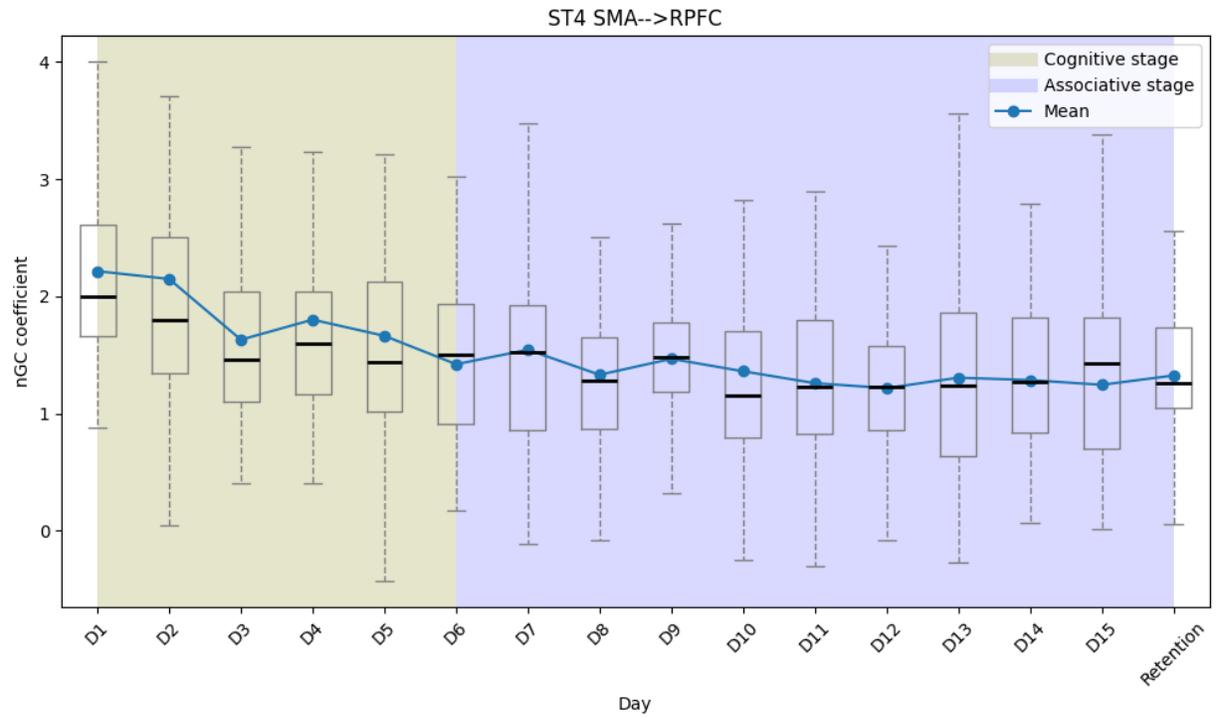

(d)

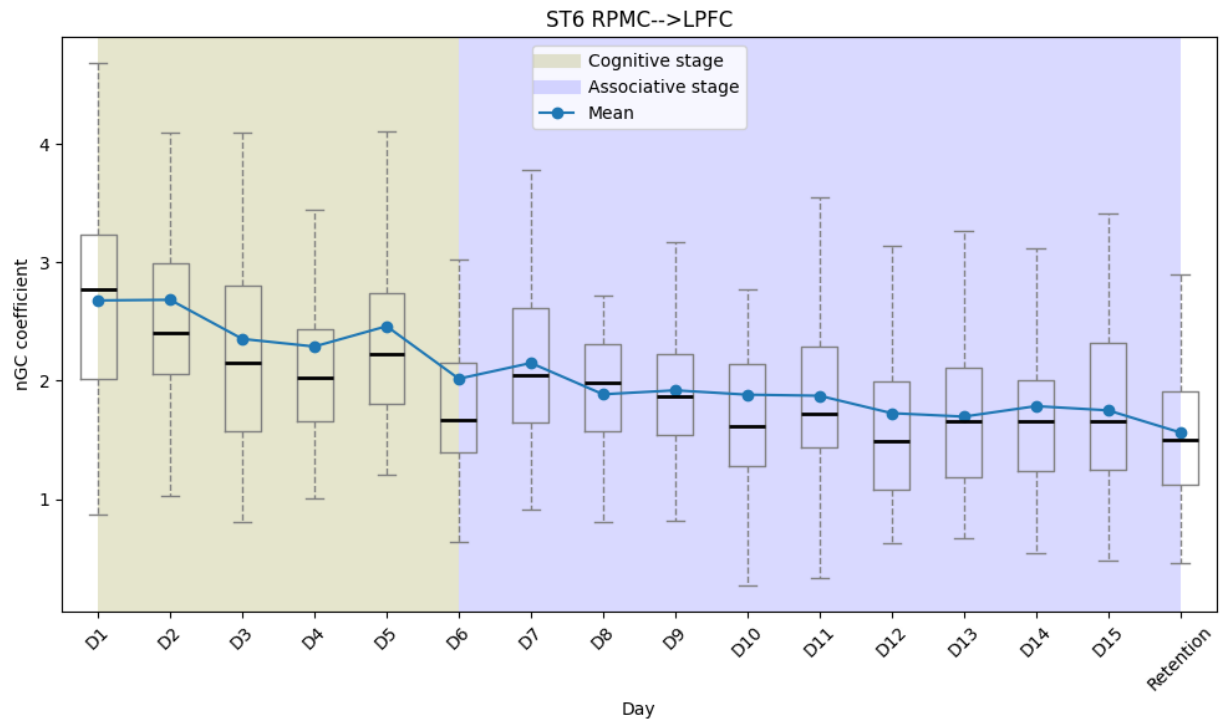

(e)

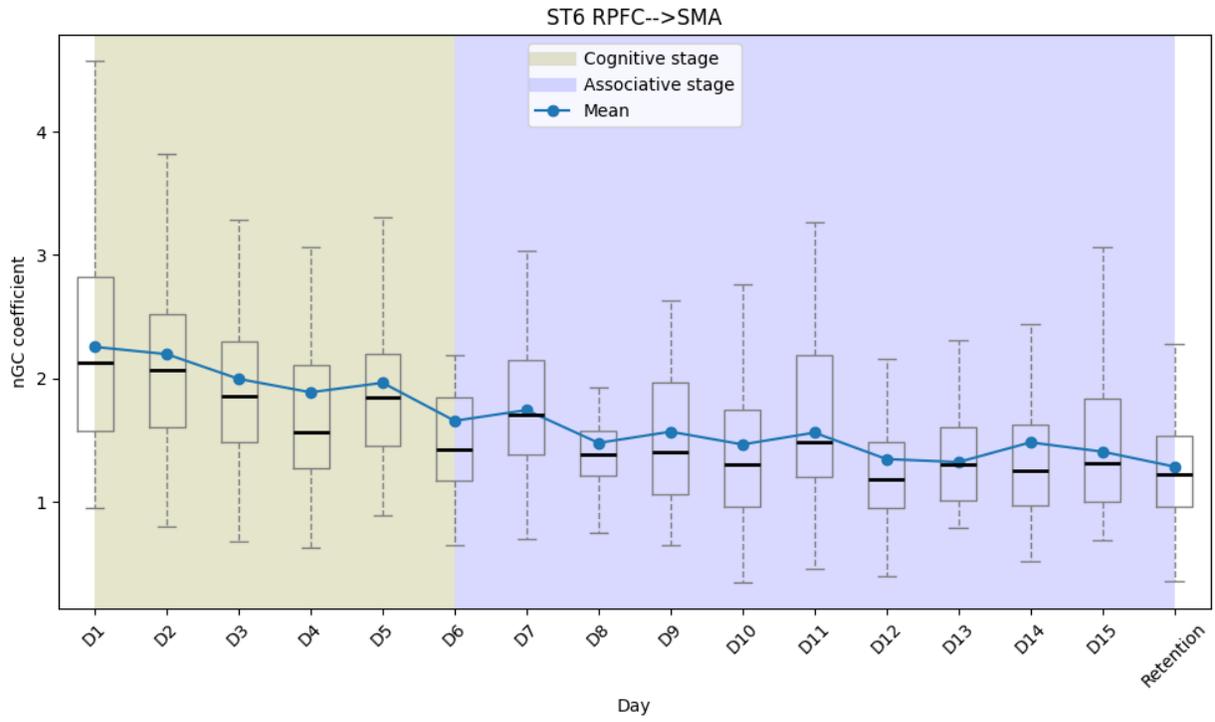

(f)

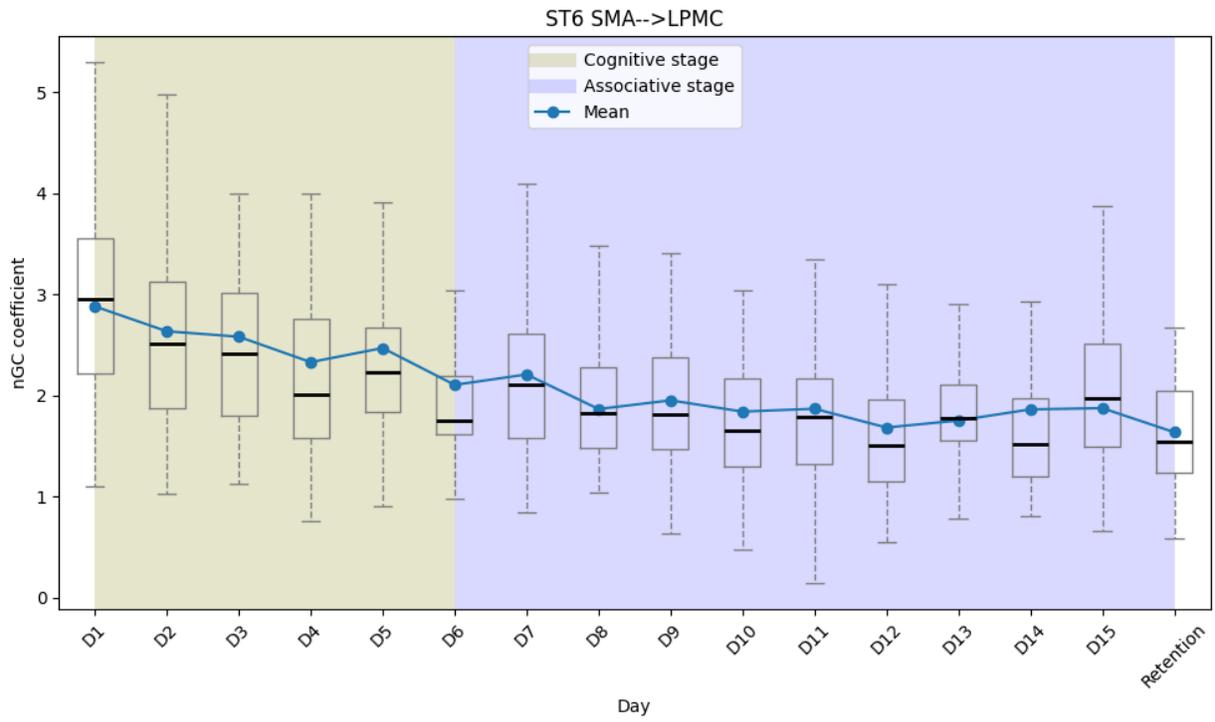

(g)

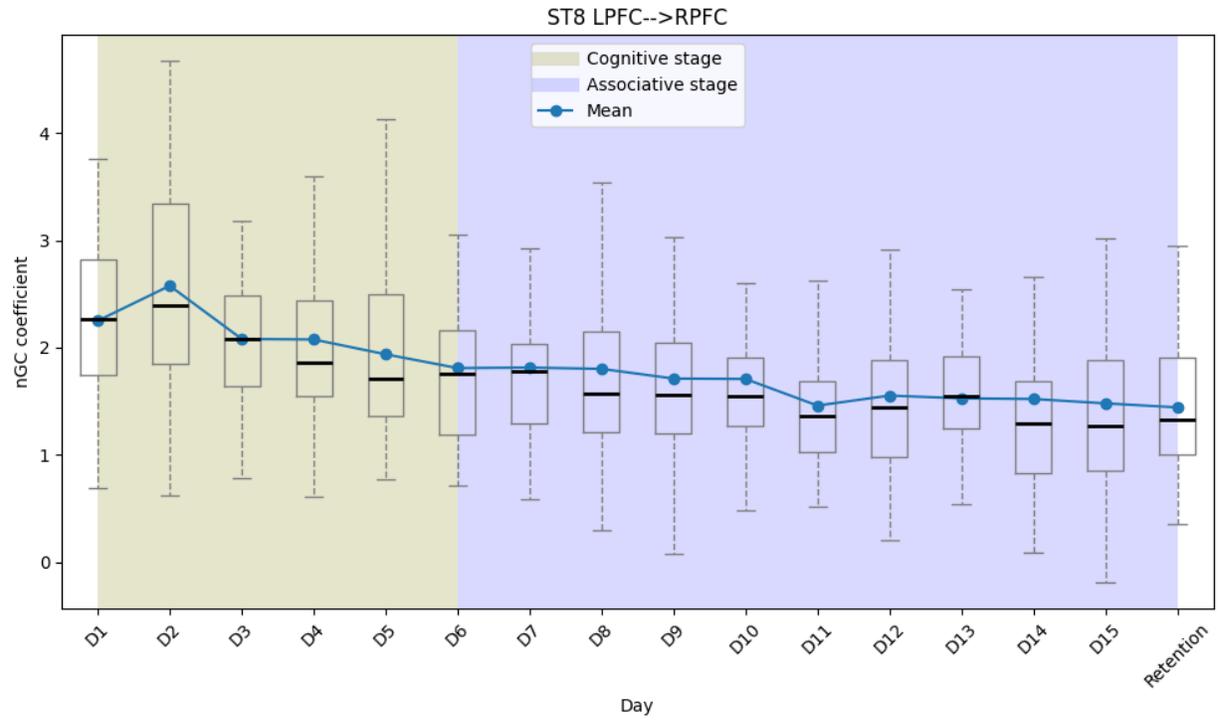

(h)

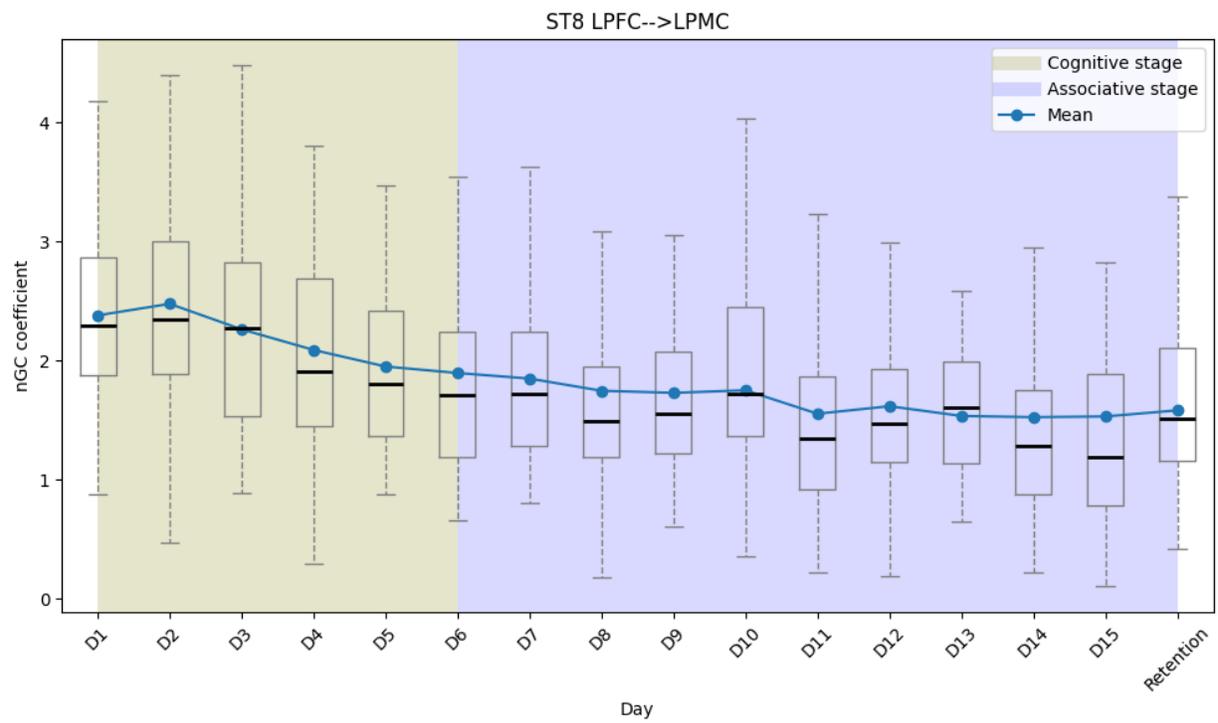

(i)

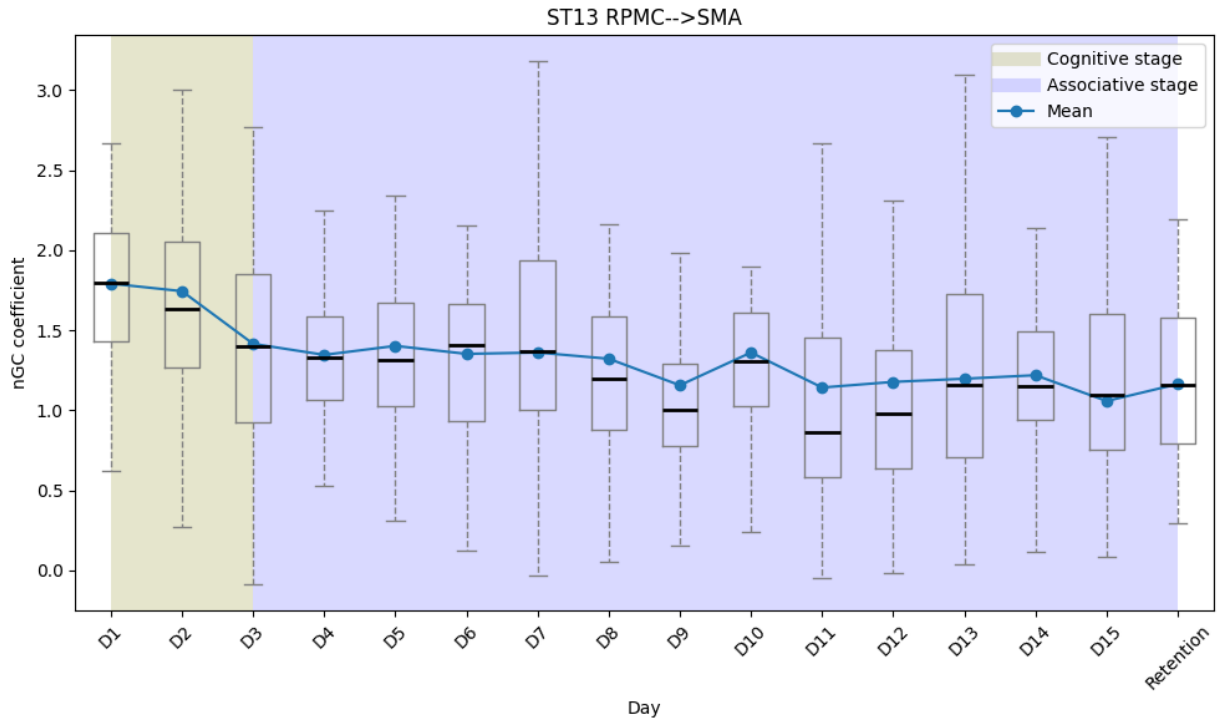

(j)

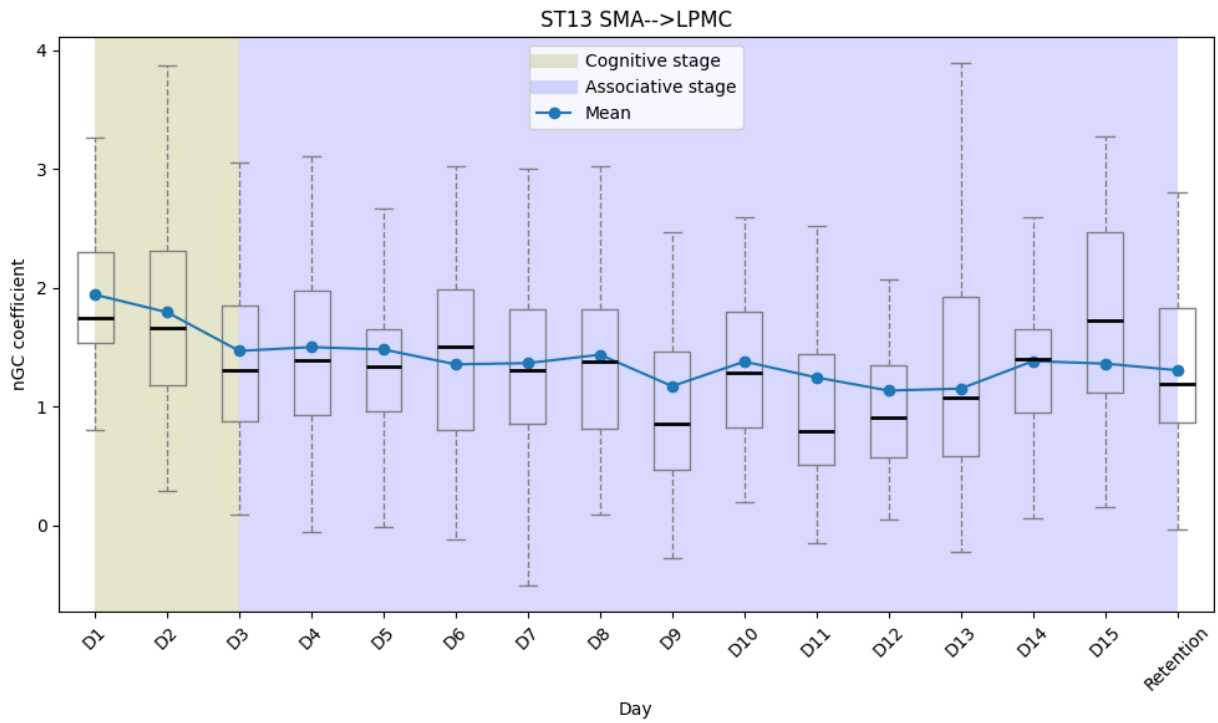

**S. Figure 2:** Strength of dominant connectivities at various adapting subtasks ((a-b) ST3, (c) ST4, (d-f) ST6, (g-h) ST8 and (i-j) ST13) over training and retention days. Mean of the dFC on each day is joined by a connecting line. The identified transition day based on dFC from cognitive to associative stage is marked by a vertical red dot line.

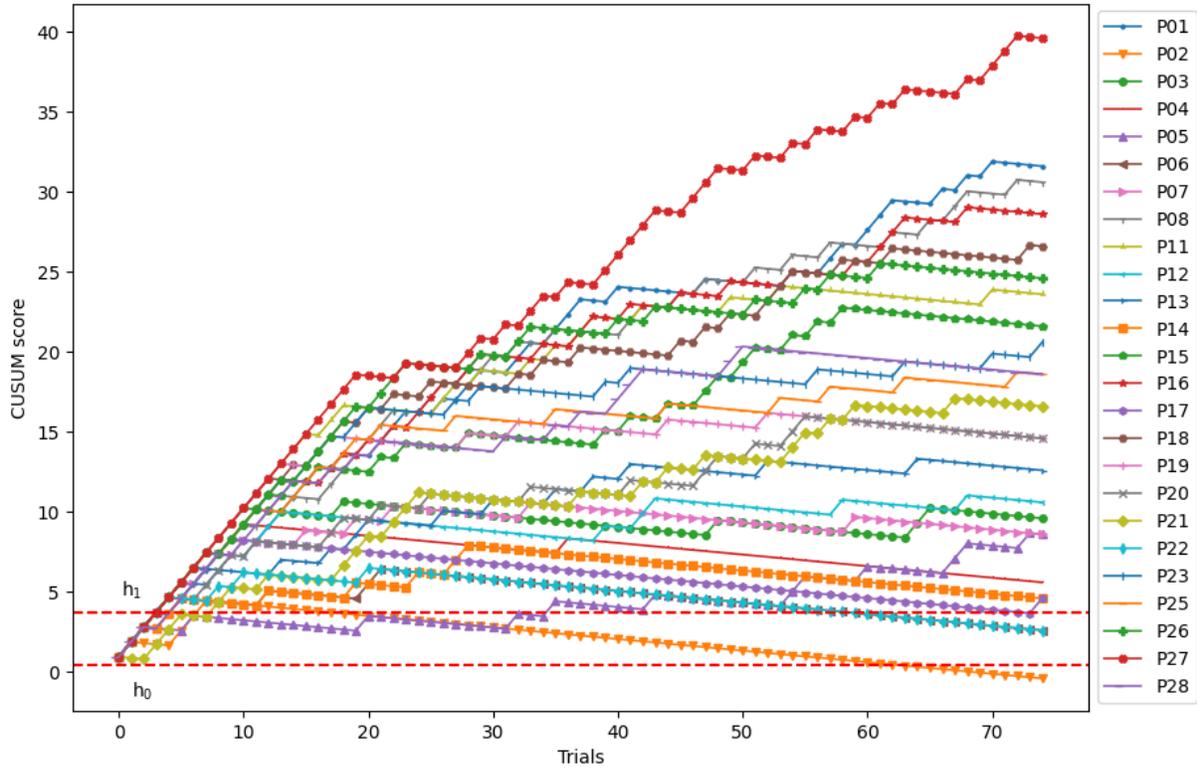

**S. Figure 3:** CUSUM plot of trainees over the training days. "$h_0$" and "$h_1$" threshold are shown by the horizontal dot lines. The trainee's code name is listed on the right.

(a)

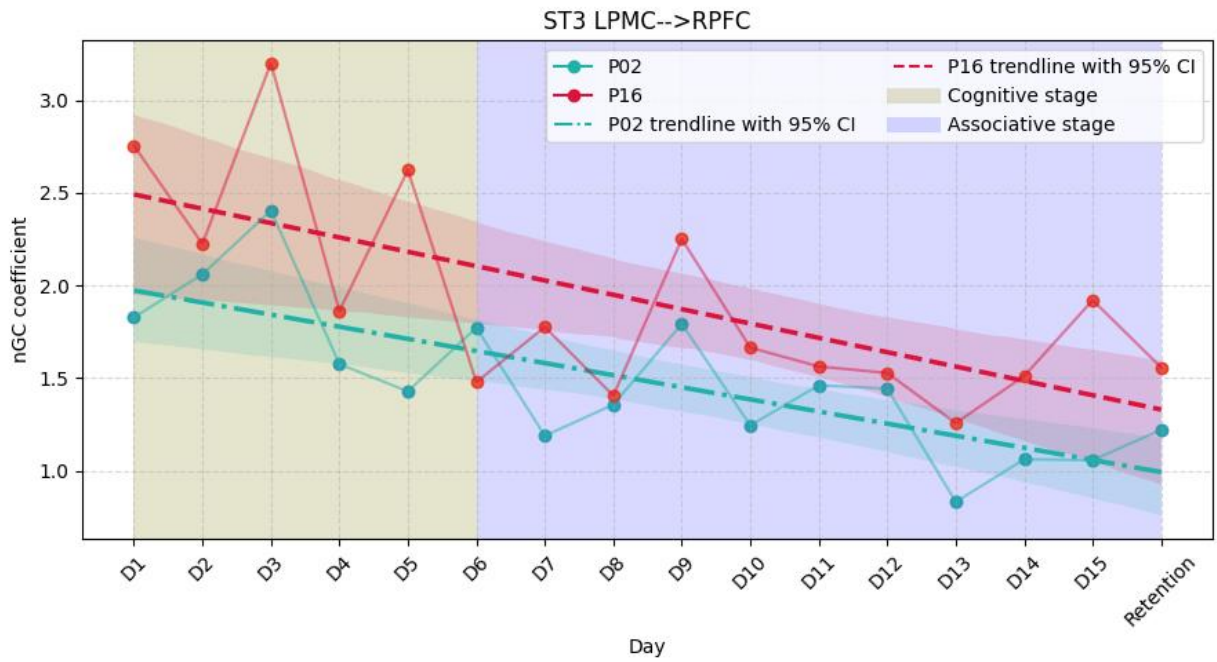

(b)

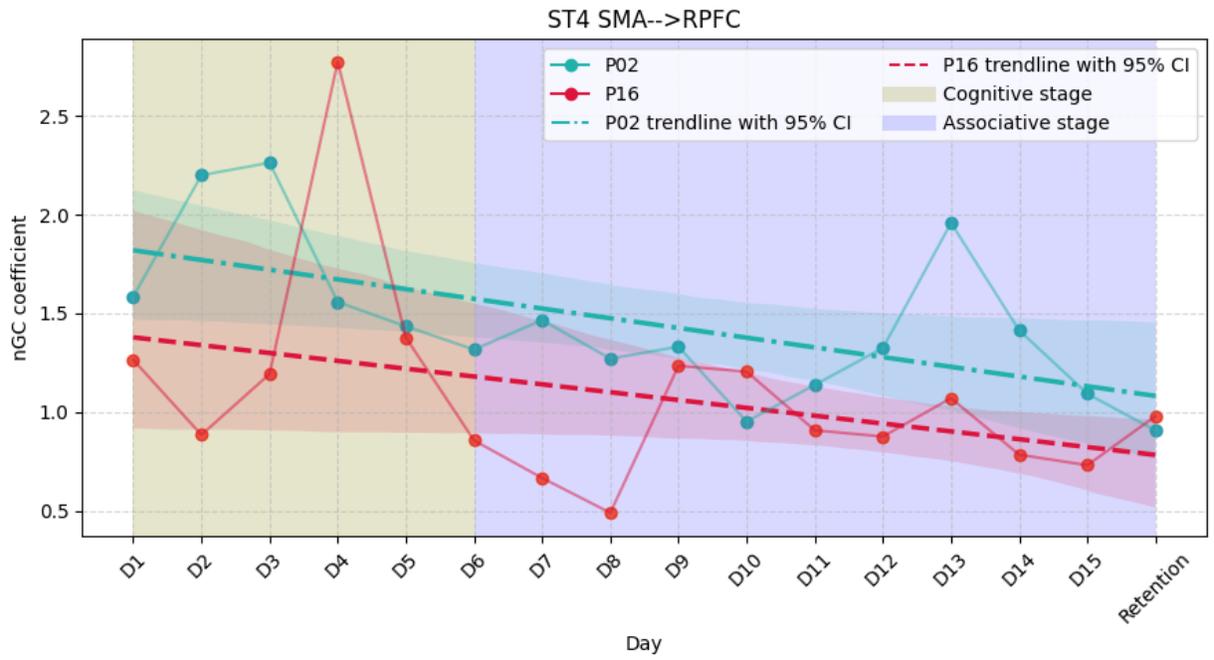

(c)

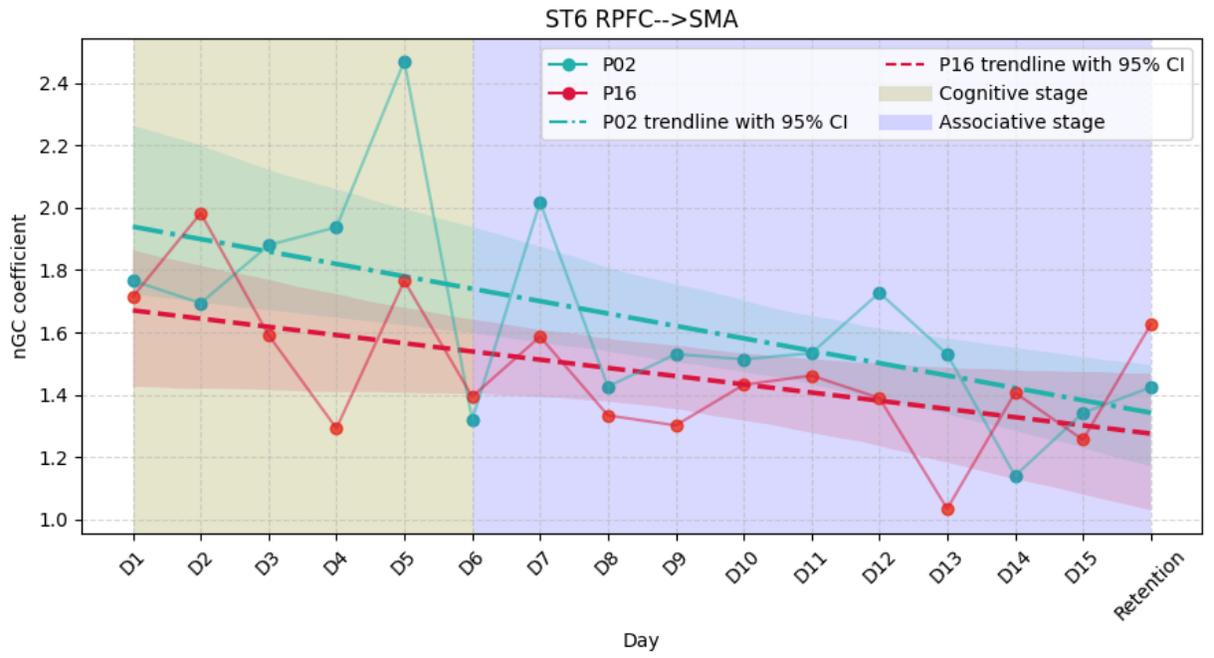

(d)

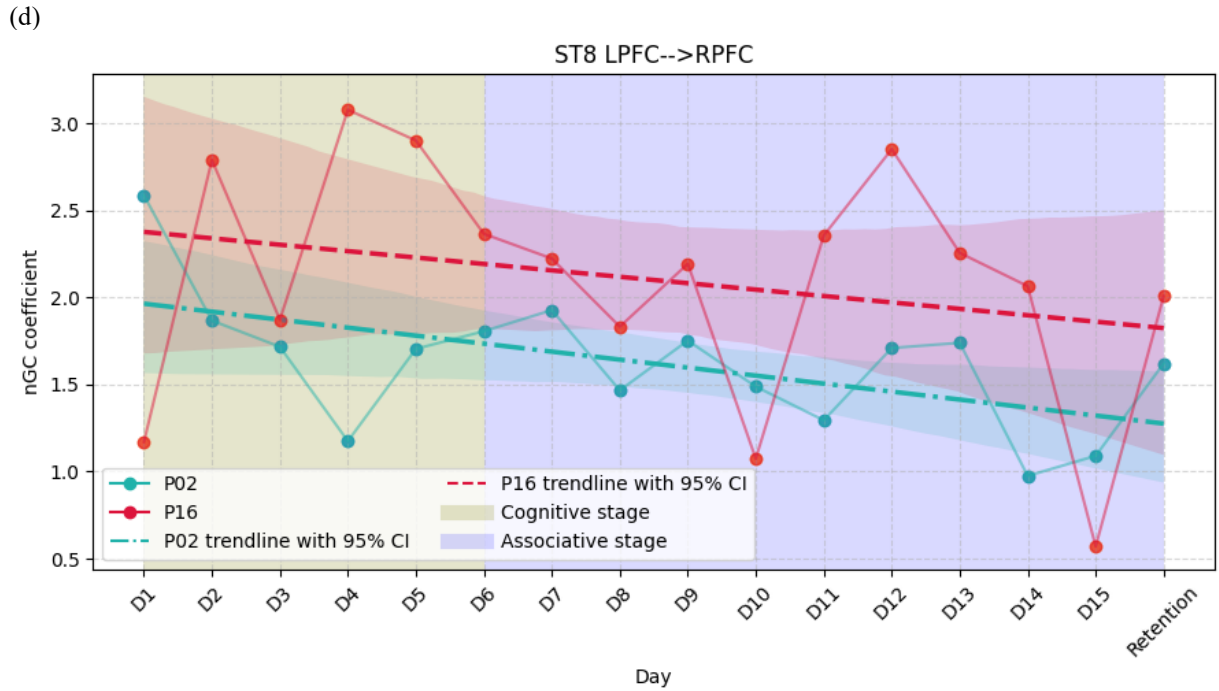

(e)

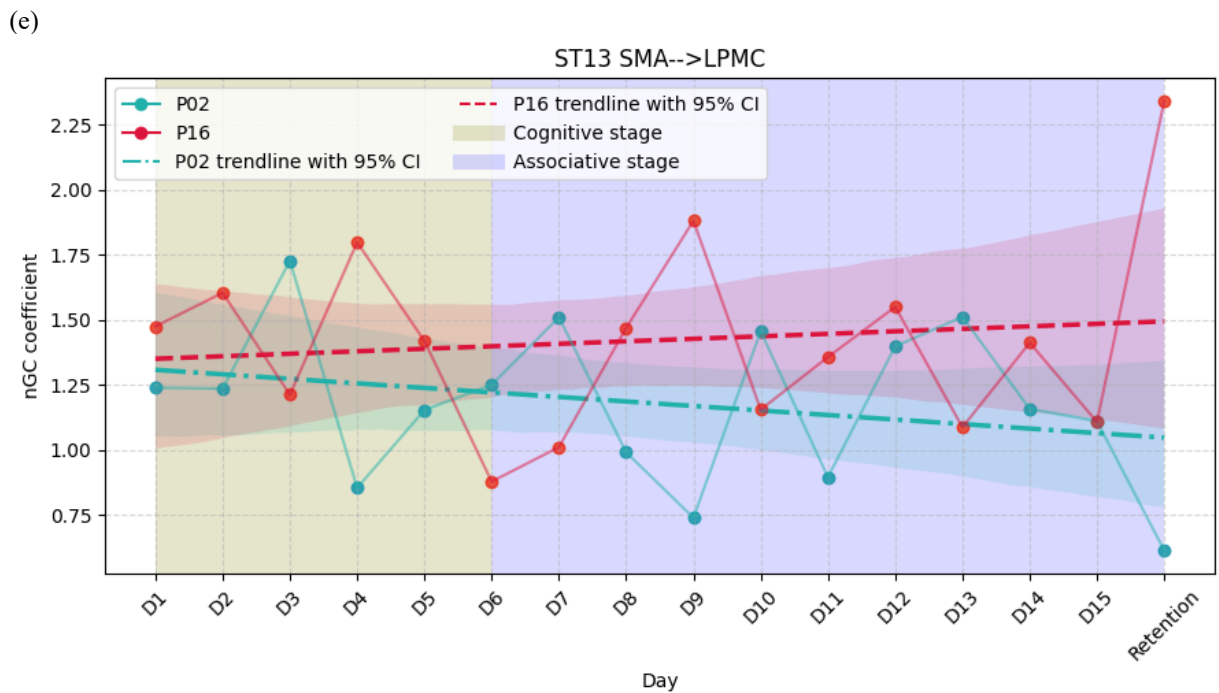

**S. Figure 4:** Evolution of dFC for individual trainees P02 (crossing"h0" threshold) and P16 (not crossing"h0" threshold) for adapting subtasks (a) ST3 (b) ST4 (c) ST6 (d) ST8 (e) ST13. Linear regression is used to show the trend in the dFC over training days for both the trainees, with 95% confidence interval (CI) shown in shaded regions around the trend line. The transparent background colors represent motor learning stages, with a transition from the cognitive to the associative stage observed on the sixth day of training based on dFC.

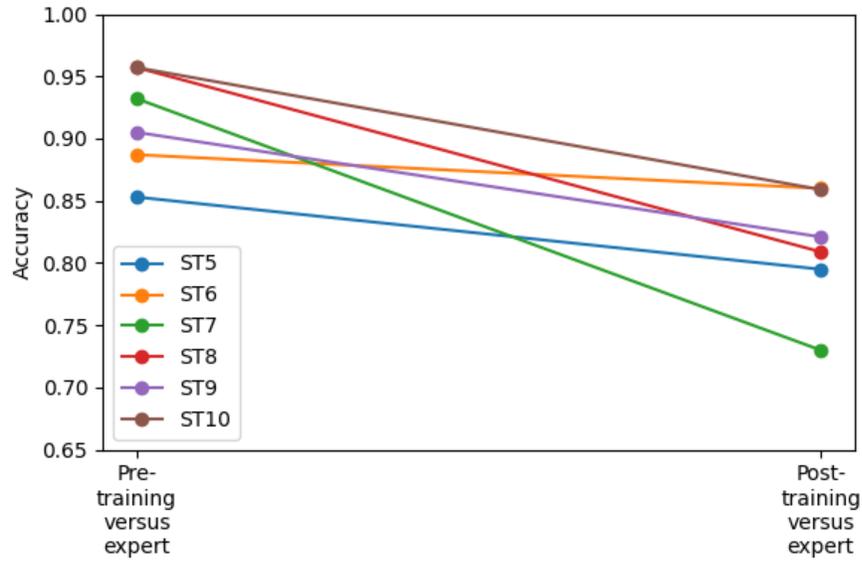

**S. Figure 5:** Comparison of pre-and post-training connectivities with expert for some of the complex subtasks.

| Subtasks | Accuracy | Sensitivity | Specificity | MCC | AUC |
| --- | --- | --- | --- | --- | --- |
| ST1 | 0.716 | 0.629 | 0.752 | 0.361 | 0.726 |
| ST2 | 0.692 | 0.452 | 0.792 | 0.247 | 0.624 |
| ST3 | 0.652 | 0.565 | 0.689 | 0.238 | 0.639 |
| ST4 | 0.719 | 0.541 | 0.792 | 0.328 | 0.671 |
| ST5 | 0.705 | 0.452 | 0.811 | 0.271 | 0.630 |
| ST6 | 0.615 | 0.467 | 0.676 | 0.134 | 0.551 |
| ST7 | 0.628 | 0.350 | 0.741 | 0.092 | 0.522 |
| ST8 | 0.624 | 0.500 | 0.676 | 0.165 | 0.598 |
| ST9 | 0.660 | 0.542 | 0.709 | 0.238 | 0.637 |
| ST10 | 0.643 | 0.390 | 0.750 | 0.141 | 0.529 |
| ST11 | 0.680 | 0.305 | 0.841 | 0.166 | 0.491 |
| ST12 | 0.653 | 0.491 | 0.721 | 0.204 | 0.608 |
| ST13 | 0.730 | 0.518 | 0.820 | 0.343 | 0.641 |

**S. Table 5:** Comparison of dFC on post-training with retention day.

(a)          (b)

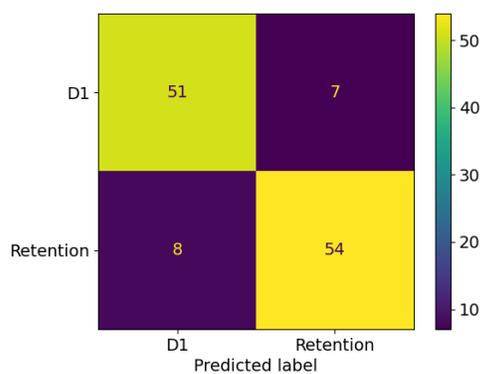

(c)

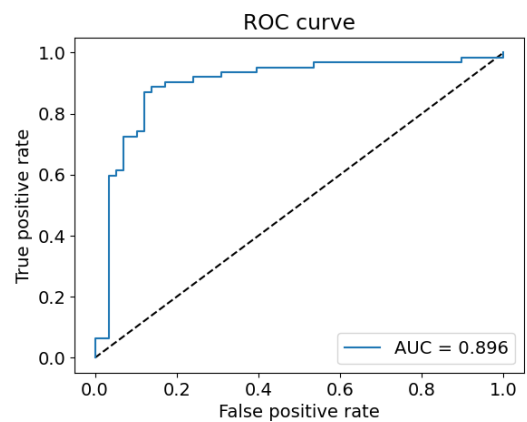

(d)

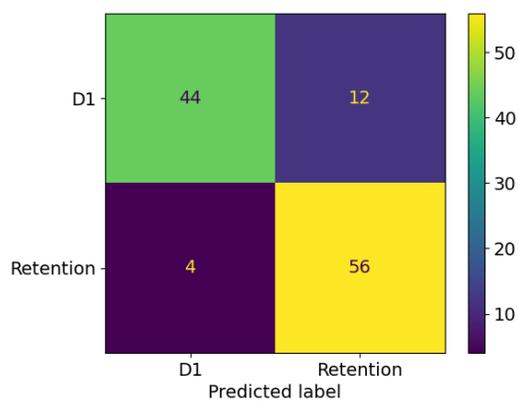

(e)

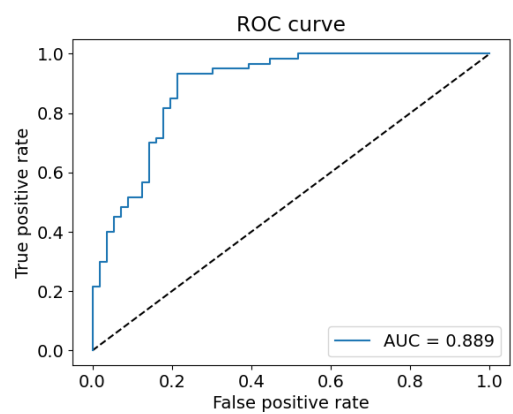

(f)

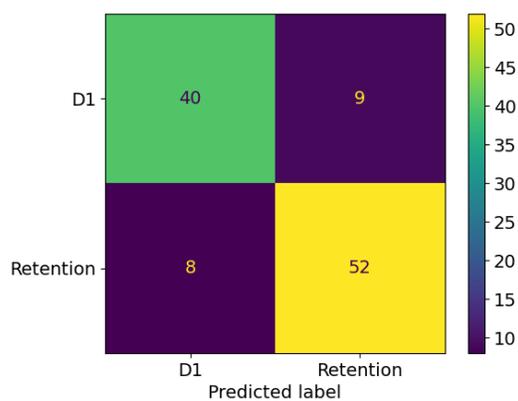

(g)

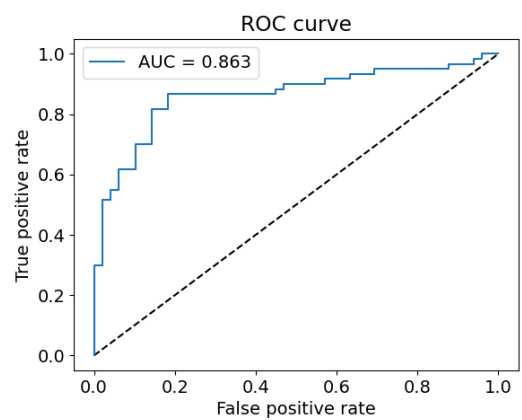

(h)

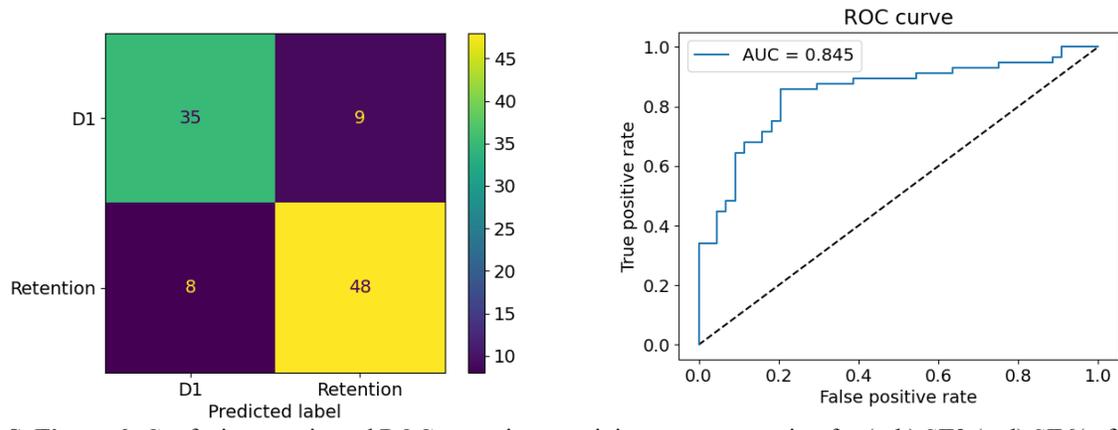

**S. Figure 6:** Confusion matrix and ROC curve in pretraining versus retention for (a-b) ST3 (c-d) ST6(e-f) ST8 (g-h) ST13.